\documentclass[journal,twoside,web]{ieeecolor}
\usepackage{generic}
\usepackage{cite}
\usepackage{amsmath,amssymb,amsfonts}
\usepackage{algorithmic}
\usepackage{graphicx}
\usepackage{algorithm,algorithmic}
\usepackage{hyperref}
\hypersetup{hidelinks=true}
\usepackage{booktabs}
\usepackage{multirow}

\def\rsq{\hspace*{\fill}$\square$\medskip}

\def\BibTeX{{\rm B\kern-.05em{\sc i\kern-.025em b}\kern-.08em
    T\kern-.1667em\lower.7ex\hbox{E}\kern-.125emX}}
\begin{document}
\title{Physics-informed DeepCT: Sinogram Wavelet Decomposition Meets Masked Diffusion}
\author{Zekun Zhou$^{1}$, Tan Liu$^{2}$, Bing Yu$^{2}$,  Yanru Gong$^{2}$,  Xi Tao$^{3}$, Liu Shi$^{2}$,  Qiegen Liu$^{2}$,~\IEEEmembership{Senior Member,~IEEE} 
\vspace{-2 em}
\thanks{This study was funded by National Natural Science Foundation of China (621220033, 62201193). }
\thanks{This study was funded by the Nanchang University Youth Talent Training Innovation Fund Project (Grant: XX202506030012).}\thanks{This study was funded by the Open Project of the Key Laboratory of Advanced Medical Imaging and Intelligent Computing of Guizhou Province (No. AMIIC250102)}
\thanks{This work was supported by data from YOFO  Medical Technology Co.,Ltd.(Hefei, China). }
\thanks{$^{1}$School of Mathematics and Computer Sciences, Nanchang University, Nanchang, China.}
\thanks{$^{2}$School of Information Engineering, Nanchang University, Nanchang, China.}
\thanks{$^{3}$Key Laboratory of Advanced Medical Imaging and Intelligent Computing of Guizhou Province, China.}
\thanks{Z. Zhou (ZekunZhou@email.ncu.edu.cn) and T. Liu are co-first authors.  Co-corresponding authors: L. Shi (shiliu@ncu.edu.cn) and Q. Liu (liuqiegen@ncu.edu.cn).}%
}

\maketitle

\begin{abstract}
Diffusion models have demonstrated strong potential in sparse-view computed tomography (SVCT) reconstruction. However, their generalization ability is often constrained when trained on limited sample spaces, leading to performance degradation when encountering unseen data. This typically leads to image blurring, loss of structural details, and cross-region inconsistencies. To address these challenges, we propose a \textbf{S}inogram-based \textbf{W}avelet random decomposition \textbf{A}nd \textbf{R}andom mask diffusion \textbf{M}odel (SWARM) for SVCT reconstruction. Specifically, introducing a random mask strategy in the sinogram effectively expands the limited training sample space. This enables the model to learn a broader range of data distributions, enhancing its understanding and generalization of data uncertainty.  In addition, we designed a wavelet-based random training mechanism for sinogram high-frequency components, enabling the model to capture structural details in different frequency bands and enhancing the richness and structural consistency of the representations. Two-stage iterative reconstruction method is adopted to ensure the global consistency of the reconstructed image while refining the details. Compared with other state-of-the-art reconstruction methods, SWARM can increase the PSNR by 3.59 dB on average, the SSIM by 0.69\%, and reduce the MSE by 55.40\%. These experimental results indicate that SWARM has great potential in the field of sparse-view CT image reconstruction.
\end{abstract}


\section{Introduction}
\label{sec:introduction}

\IEEEPARstart{S}{parse-view} X-ray computed tomography (SVCT) is extensively studied for its low-dose and rapid imaging benefits in medical diagnosis and industrial non-destructive testing \cite{cormack1963representation,hounsfield1973computerized}.  However, due to the incomplete of data acquisition of SVCT, serious artifacts are introduced into the reconstructed images, obscuring important internal structures and features\cite{Wang2024A}. SVCT reconstruction is a challenging inverse problem and improving the reconstruction quality has been a frontier in recent years\cite{Li2023Sparse-view}. 

Classical iterative reconstruction methods\cite{dempster1977maximum} were proposed to improve the image quality though the performance was still poor on highly sparse views. Compressed sensing\cite{donoho2006compressed} utilized priors applicable to sparse data such as total variation (TV)\cite{Sidky2008Image , Yu2009Compressed}  and wavelet frame\cite{rioul1991wavelets}, demonstrating a powerful ability to data with sparsity. However, these methods are computationally expensive due to the iterative update steps required and the effect is limited by the parameter sensitivity\cite{Han2016Deep , Zhang2018A}.

Deep learning-based reconstruction models have gained increasing attention for their high computational speed\cite{Koetzier2023Deep}  and robustness of parameter\cite{Zhong2024Impacts}. For example, post-processing methods for CT image domain reconstruction include FBPConvNet\cite{McCann2017Deep}, dense deconvolution networks\cite{Zhang2018A}, and residual encoder-decoder convolutional neural networks\cite{Chen2017Low-Dose}. Additionally, hybrid domain processing methods have been proposed to reconstruct high-quality images by learning from both the projection domain and the image domain\cite{Zhang2020Artifact, Pan2022Multi-domain}, such as hybrid domain neural network\cite{Hu2021Hybrid-Domain}. However, the performance of these methods for image reconstruction is still limited.

In recent years, diffusion models have been increasingly employed in SVCT reconstruction owing to their superior  capabilities.  Xia \textit{et al.}\cite{Xia2022Patch-Based} introduced a patch-based denoising probabilistic diffusion model to improve SVCT reconstruction performance and address large memory requirements. Additionally, Wu \textit{et al.}\cite{Wu2023Data-iterative} proposed an iteratively optimized data scoring model grounded in SGM to achieve high-quality CT reconstruction for ultra-sparse views. Guan \textit{et al.}\cite{Guan2023Generative} introduced a score-based diffusion model that uses a multi-channel strategy in the projection domain to ensuring that the generated information closer to the original data, resulting in  a more accurate SVCT reconstruction. Xu \textit{et al.}\cite{10403850} used a diffusion model for stage-by-stage optimization in the wavelet domain, which  enhanced sparse-view CT image quality. Xia \textit{et al.}\cite{Xia2023Parallel} converted the denoising diffusion probability model into a parallel framework to improve the efficiency  of the model, and applied it to the reconstruction of breast CT images in dual-domain sparse view. Yang \textit{et al.}\cite{Yang2024A} introduced a dual-domain diffusion model  for SVCT reconstruction, which includes a sinogram enhancement module and an image refinement module. Although diffusion models have achieved some success in SVCT reconstruction, they still have certain limitations in capturing finer information\cite{kazerouni2023diffusionmodelsmedicalimage} and rely on large amounts of high-quality data for training\cite{ho2020denoisingdiffusionprobabilisticmodels}.

To partially address these challenges, masked diffusion encourages the model to attend to critical regions or structural features, thereby enhancing its ability to reconstruct local details in data-scarce settings.  Moreover, introducing perturbations into the input data can enhance the robustness of the model against missing information and improve the generalization performance of the model. Aversa \textit{et al.}\cite{aversa2024diffinfinite} used a layered diffusion model to generate synthetic segmentation masks for high-fidelity diffusion, reducing reliance on labeled data and preserving image detail and structure. Toker \textit{et al.}\cite{toker2024satsynthaugmentingimagemaskpairs} used a denoising diffusion probability model to simultaneously generate images and masks, addressing data scarcity in satellite segmentation tasks while ensuring a wide diversity of samples. Konz \textit{et al.}\cite{konz2024anatomicallycontrollablemedicalimagegeneration} applied masked diffusion to medical image segmentation, demonstrating its advantages when dealing with complex anatomical structures and reducing the reliance on large amounts of high-quality training data. With the successful application of masked diffusion, it has been gradually introduced into the field of SVCT reconstruction. For example,  Tan \textit{et al.}\cite{tan2024msdiff} proposed a score-based multiscale diffusion model. By introducing a regular mask into the SVCT reconstruction framework, this approach effectively  improved model performance. Nevertheless, the reliance on fixed masking strategies imposes constraints on the network's capacity to learn diverse data distributions, thereby limiting the generalization of the model.

To address the above issues, we introduce a \textbf{S}inogram \textbf{W}avelet random decomposition \textbf{A}nd \textbf{R}andom mask  diffusion \textbf{M}odel (SWARM). The proposed random mask strategy effectively expands the limited training sample space, allowing the model to learn from a broader range of data distributions. This dual mechanism not only reinforces the model's capacity to characterize epistemic uncertainty but also substantially elevates its generalization capability. By integrating stochastic masking into sinogram projections, our approach fosters a synergistic learning framework that enables the model to encode global structural dependencies while enhancing its robustness in extrapolating to unseen data distributions. To further refine reconstruction quality, we apply wavelet-based random high-frequency decomposition, which improves the representation of fine structural details across frequency bands. Finally, a two-stage iterative reconstruction framework with data consistency constraints ensures both global structural coherence and accurate local detail recovery.

The main contributions of this paper can be summarized as follows:

\indent $\bullet$ \textbf{Multi-scale Joint Global-Detail Dual Diffusion.} We propose a global-detail integrated training framework that incorporates a dual diffusion model.  By harnessing multi-scale analysis, this methodology effectively captures both global contextual information and fine-grained features in CT reconstruction tasks, thereby significantly enhancing the model's comprehension of image structural hierarchies.

\indent $\bullet$  \textbf{Uncertainty Increasing of Random Mask Embedding in Sinogram Training.}  We introduce a strategy of embedding sinogram with random masks to simulate incomplete sampling scenarios, thereby augmenting uncertainty and diversity in the training process.  This approach induces feature-space uncertainty, enriches training data diversity, and enhances model generalization capability as well as reconstruction robustness under distribution shifts.

\indent $\bullet$  \textbf{Robust Feature Learning for Sinogram  Random High-frequency.}  The sinogram is decomposed into multi-frequency subbands via wavelet transform, with high-frequency components undergoing random sampling to serve as training inputs. This frequency-sensitive sampling strategy enhances spatial feature discriminability and fortifies the model’s robustness against structural variations and noise perturbations.

In Section II, we provide a brief review of the relevant work.  Section III outlines the theoretical approach and offers a detailed explanation of our proposed method.  Section IV presents the experimental comparison results.  Finally, in Section V, we discuss and conclude the methods presented.

\section{Preliminary}

\subsection{Sparse-view CT Image Reconstruction}

Given an original image $x$, it represents the linear attenuation coefficient distribution of internal structures within the object, reflecting the differential absorption characteristics of heterogeneous tissues to X-ray photons. The image $x$  is converted into full projection data through radon transformation, which is expressed as $R(\theta,s)= \mathfrak{R}(x)(\theta,s)$. $R(\theta,s)$  denotes the outcome of the radon transformation at the position s for angle $\theta$  along the projection direction.  $\mathfrak{R}(x)(\theta,s)$  is the radon transform operator. The projection data obtained through scanning is a linear mapping of noise observation data to a set of measured values, expressed as:

\begin{equation}
\label{eq1}
y= Ax+  \varepsilon, 
\end{equation}
where  $y$ represents the measured projected data, $A$  is the system matrix determined by the geometry of CT equipment and the scanning protocol used and $\varepsilon$ represents the system error and random noise.  SVCT reconstruction is a classical inverse problem \cite{song2021solving}. Owing to the incompleteness in the data acquisition process, accurately reconstructing unknown images from limited measurements presents a significant challenge. The mapping from full projection data  $y$ to sparse-view projection data $\hat{y}$  is a process of linear transformation. Specifically, for a linear mapping function $f:\mathbb{R}^{m\times n} \rightarrow \mathbb{R}^{m\times n} $, acting on $y$  by the linear operator $P(\wedge)$,   is obtained and expressed as: 

\begin{equation}
\label{eq2}
\hat{y}=f(y)=P(\wedge)y.   
\end{equation}

Using the traditional FBP algorithm to reconstruct images directly from SVCT projection data can lead to severe fringe artifacts and blurred details.  To improve the quality of reconstruction, it is usually necessary to incorporate prior information into the regularized objective function  to address the following issues:

\begin{equation}
\label{eq3}
y^{\ast}=\underset{y}{\arg\min} \ \frac{1}{2} \lVert P(\wedge)y-\hat{y}  \rVert ^{2} _{2} + \frac{\nu}{2}R(y), 
\end{equation}
where the first item is the data fidelity to ensure that the actual data obtained from under-sampling is aligned with the obtained measurement values. The second term is the regularization term, and the hyperparameter  $\nu$ used to balance the data fidelity term and the regularization term.

\subsection{Masked Diffusion}
{

Diffusion models have achieved remarkable success in image, audio, and text fields with their high-quality data generation capability. In the forward process, the model gradually transforms the initial data distribution into a gaussian distribution by progressively adding noise\cite{acar2024advanced,ozturk2024diffusion,ozbey2023unsupervised,gungor2023adaptive}.  In the reverse process, the model learns a reverse denoising process to gradually remove the noise, ultimately achieving accurate recovery of the original data. The research on diffusion model mainly focuses on the following aspects: denoising diffusion probabilistic models (DDPMs) \cite{ho2020denoisingdiffusionprobabilisticmodels}, score-based generative models (SGMs)\cite{song2019generative}, and stochastic differential equations(SDEs)\cite{song2020score}.

Masked diffusion has recently emerged as a powerful technique in image processing, demonstrating strong capabilities in learning data representations from incomplete inputs. By applying binary masks to input images, masked diffusion enables models to reconstruct missing regions, thus enforcing semantic understanding and structural reasoning. This masking strategy encourages the model to focus on global coherence while generating local details, which has proven effective in self-supervised and unsupervised settings alike.

While existing works have leveraged masked diffusion in diverse tasks such as image synthesis\cite{wang2024flame,gao2023masked,aversa2024diffinfinite}, image editing\cite{couairon2022diffedit,wang2023instructedit,zou2024towards}, image restoration\cite{pang2024improved,zhu2023denoising}, image segmentation\cite{le2024maskdiff,toker2024satsynth}. Most approaches focus on natural images and overlook the unique challenges of projection data in medical imaging. In particular, conventional masking strategies fail to simulate the complex patterns of missing or corrupted sinogram data encountered in SVCT reconstruction. To address this gap, by introducing random occlusion in the projection domain, our method enhances the model's ability to generalize from finite and degenerate inputs.

}

\section{Method}

\subsection{Motivation}

In medical imaging, reconstructing high-quality images from sparse-view projections is vital for accurate diagnosis. Although traditional deep learning models achieve satisfactory performance when trained and tested within closed datasets. Their generalization ability degrades significantly when deployed on out-of-distribution (OOD) or clinically diverse data.  Abdar et al. \cite{ABDAR2021243} and Huang et al. \cite{HUANG2025103334} emphasize the risk of overfitting to training distributions and the consequent drop in diagnostic reliability. However, current literature rarely explores effective strategies to explicitly enhance robustness to distribution shifts in the projection domain, especially in the context of SVCT. Most approaches focus on either improving reconstruction architectures or increasing data diversity through simple augmentation, failing to systematically address the sensitivity of models to the variability in sampling patterns and acquisition conditions.

\begin{figure}[!t]
\centerline{\includegraphics[width=0.97\columnwidth]{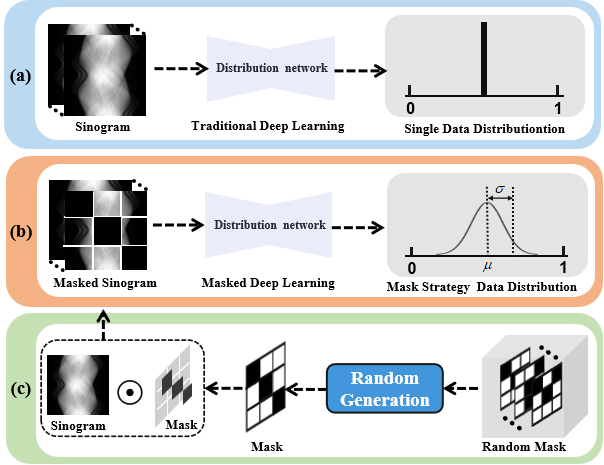}}
\caption{Different training strategies and influence of sinogram in deep learning. (a) Distribution of training data in a closed data space; (b) The distribution of the data obtained through the mask extension method in the extended data space; (c) The generation process of the random mask and how it is embedded in the data.}
\vspace{-1 em}
\label{mask}
\end{figure}

To address this gap, we propose a novel training strategy that embeds structured randomness into the projection data via random masks. By perturbing full-view sinograms during training, we simulate diverse acquisition scenarios, encouraging the model to learn robust, generalizable reconstruction mappings. As illustrated in Fig. \ref{mask}(b)-(c), the embedded random masks induce broader and more uncertain data distributions, which can improve the model's robustness to unseen clinical variations.

To theoretically understand how perturbed data influences data diversity generation mechanisms, this paper constructs a theoretical interpretive framework and systematically analyzes full-view projection data and the random masking of such data.

\noindent \textbf{Proposition 3.1:}
The incorporation of random masks into a finite sinogram sample set $y=\{ y_{1},y_{2}, \cdots , y_{n} \}$, serves to augment the variance of the data distribution.

See Appendix II for proof.

\noindent \textbf{Proposition 3.2:}
Let the original data sample space be \( y = \{y_1, y_2, \dots, y_n\} \), and \( \tilde{y} = y + m \odot y \) denote the masked sample space, where  \( m_i \sim  \mathnormal{U}(0,1) \).  The covariance of the masked data satisfies \( \tilde{\Sigma} \geq \Sigma \).  Consequently, for any direction \( \nu \in \mathbb{R}^d \), it holds that \( \nu^T \tilde{\Sigma} \nu \geq \nu^T \Sigma \nu \), which may lead to the expansion of the data distribution in the masked data space.

See Appendix III for proof.

Our analysis indicates that perturbed data increase the variance of the dataset, thereby expanding the distribution that the model learns from. This form of uncertainty-driven data perturbation helps enhance the model’s generalization ability. It is worth noting that the benefit of increased variance depends on the compatibility between the perturbed data distribution and the target test distribution. Therefore, our variance-based formulation should be regarded as a heuristic perspective. Subsequent experiments demonstrate that this controlled randomness improves the generalization capability in the SVCT reconstruction task.

Additionally, a two-stage iterative optimization framework is employed to jointly ensure global structural coherence and accurate high-frequency detail reconstruction. This dual-phase strategy balances fidelity and texture synthesis, enabling robust and high-quality reconstructions across diverse imaging scenarios.

\begin{figure*}[!t]
\centerline{\includegraphics[width=0.95\textwidth]{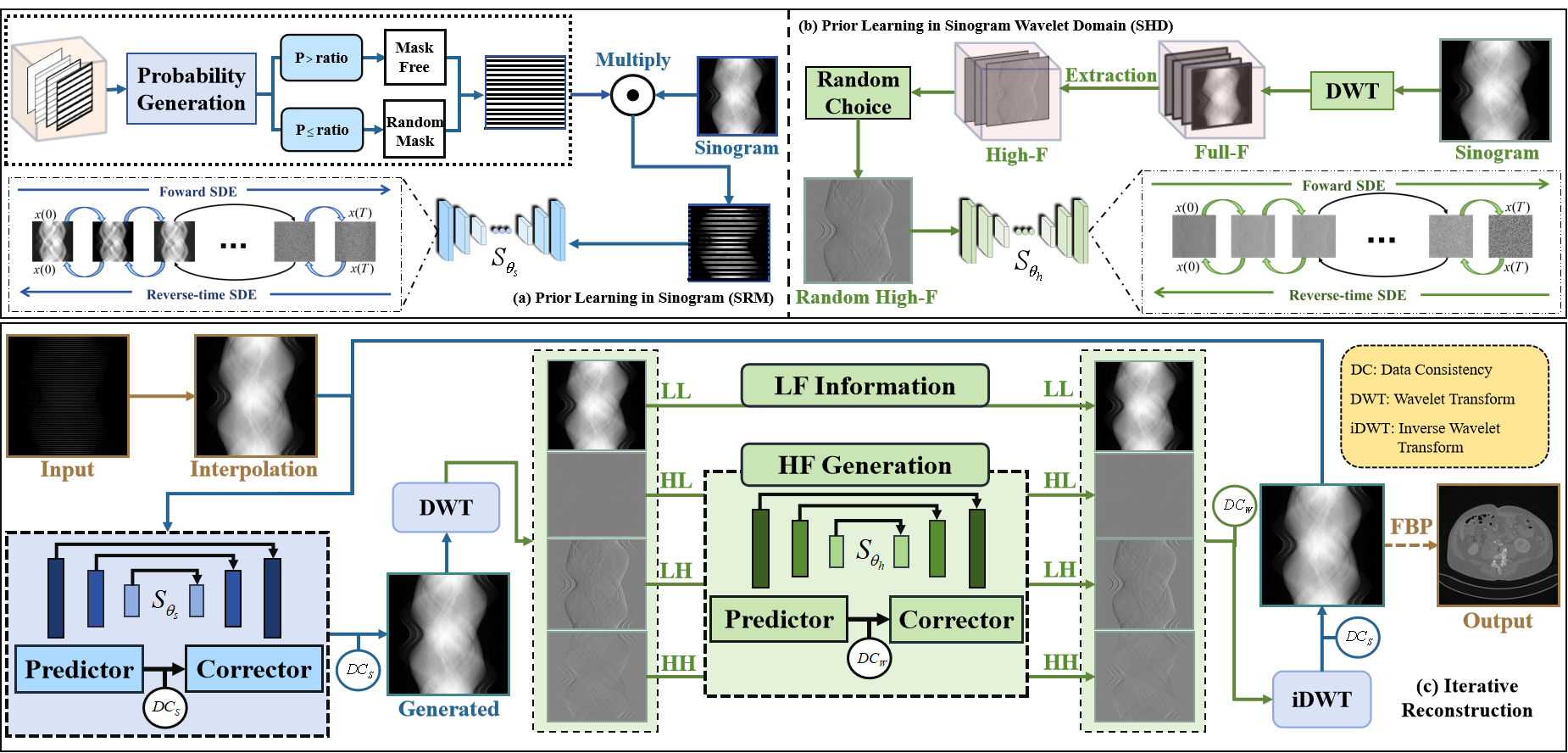}} 
\caption{The pipeline of SWARM training process and iterative reconstruction procedure. Training stage: (a) A model training based on random masks in sinogram. (b) A  model training for high-frequency random decomposition of wavelet based on sinogram. Iteration reconstruction stage: (c) The proposed SWARM method is used to reconstruct the sparse-view CT projection domain. ``LF'': Low-frequency. ``HF'': High-frequency. }
\vspace{-1 em}
\label{train+test}
\end{figure*}

\subsection{Training Process in Sinogram Wavelet Domain}

\subsubsection{Prior Learning for Random Mask Strategy in Sinogram Domain}

In the prior learning stage based on the projection domain, we propose a virtual random mask training strategy. Combining the physical characteristics of projection data acquisition and the randomness of embedding masks in the projection domain.  The \textbf{S}inogram \textbf{R}andom \textbf{M}ask model (SRM) is constructed through the SDEs to enhance the robustness of the model against distribution offsets. As shown in Fig. \ref{train+test}(a), by applying random masks to the sinograms, SRM introduces random perturbations within the limited data sample space. It effectively augments the diversity of the feature space and sharpens the model focus on a broader spectrum of data distributions. This improves the generalization ability of the model as well as the reliability and accuracy of the reconstruction results.

 By defining a probability parameter $p\in[0,1]$, this strategy is applied to simulate sparse-view sampling scenarios through stochastic masking of projection data.  Specifically, for each training sample: a sparse-view mask is applied with probability $p$.  In this case, one masking pattern is randomly selected from a predefined set of sparse-view configurations.  Given a full-view projection data $y$ and the random mask operator $\hat{m}$, the masked full-view projection data can be expressed as:
\begin{equation}
\tilde{Y}_{s}= y\odot \hat{m},  \label{eq8}
\end{equation}
where $\tilde{Y}_{s}$  indicates the projected data with the mask.  This probabilistic masking mechanism introduces controlled randomness, enhancing the model's robustness to varying degrees of data sparsity and promoting generalization across different undersampling scenarios.

The forward stochastic differential equation (SDE) progressively transforms a complex data distribution into a tractable gaussian distribution by gradually adding noise. Given a diffusion process $\{ y(t)\}^{T}_{t=0}$, represented by a continuous time variable $t \in [0,T]$, such that $y(t) \sim p_{t}$, where the sample image data is independently and identically distributed. The general forward process of SDE is expressed as follows\cite{song2021solving}:
\begin{equation}
dx= f(y,t)dt + g(t)dw,  \label{eq9}
\end{equation}
where $f(y,t):\mathbb{R}^{N} \rightarrow \mathbb{R}^{N}$  is a vector-valued function and  $\mathbb{R} \rightarrow \mathbb{R}$ is a scalar function about $y(t)$, which are called drift coefficient and diffusion coefficient respectively.  $N$ indicates the dimension of the image data in the projection domain. $w \in \mathbb{R}^{N}$   induces the standard Brownian motion. Variance explosion SDE (VE-SDE) is obtained by applying $f(y,t)=0$, $g(t)=\sqrt{d[\sigma^{2}(t)]/dt}$    in order to improve the capability.

In this case, the VE-SDE is represented as follows:
\begin{equation}
d\tilde{Y}_{s}= \sqrt{d[\sigma_{s}^{2}(t)]/dt} dw_{s},   \label{eq10}
\end{equation}
where $\sigma_{s}(t)>0$  represents a monotonically increasing function and signifies the time-varying escalating scale function for noise.

During the  prior learning stage in the projection domain, we progressively introduce gaussian noise into $\tilde{Y}_{s}$. The network $\textbf{s}_{\theta_{s}}(\tilde{Y}_{s}(t),t)$ is optimized by tuning the parameter $\theta_{s}$ to achieve optimal performance. The specific optimization objective function is given by: 
\begin{equation}
\begin{split}
\theta^{\ast}_{s} = \underset{\theta_{s}}{\arg\min} \ \mathbb{E}_{t} \{ \lambda_{t} \mathbb{E}_{\tilde{Y}_{s}(0)}  \mathbb{E}_{\tilde{Y}_{s}(t)| \tilde{Y}_{s}(0)} [ \| \textbf{s}_{\theta_{s}}(\tilde{Y}_{s}(t),t)  \\
 -\nabla_{\tilde{Y}_{s}(t)} \log p_{0t}(\tilde{Y}_{s}(t) | \tilde{Y}_{s}(0) )   \| ^{2} _{2} ] \},   
\end{split}
  \label{eq11}
\end{equation}
where $\lambda_{t}$ is a positive function, $\log p_{0t}(\tilde{Y}_{s}(t) | \tilde{Y}_{s}(0) )$ is the gaussian perturbation kernel centered at $\tilde{Y}_{s}(0)$. Once the network satisfies $\textbf{s}_{\theta_{s}}(\tilde{Y}_{s}(t),t)$  and $\nabla_{\tilde{Y}_{s}(t)} \log p_{0t}$ will be known for all $t$ by solving $\nabla_{\tilde{Y}_{s}(t)} \log p_{0t}$.

\subsubsection{Prior Learning for Random High-frequency in Sinogram Wavelet Domain}

In order to improve the detail quality of the reconstructed image, a random high-frequency training strategy based on sinogram wavelets is proposed to obtain the \textbf{S}inogram wavelet random \textbf{H}igh-frequency  \textbf{D}ecomposition model (SHD).  In this approach, the projection data is decomposed into frequency bands using orthogonal wavelet basis functions.  A dynamic random sampling mechanism is then formulated for high-frequency components, where learnable weights are employed to quantify the salience of each sub-band.   By leveraging learnable weights, the significance of each sub-band is assessed, enabling the generation of training data with anisotropic perturbations in the frequency domain. Furthermore, this method clearly distinguishes the high-frequency local details in the projection domain. This forces the network to adaptively focus on the information of key frequency bands.

Specifically, decomposing the sinogram into four subbands using wavelet transform (DWT): the low-frequency approximation component (LL) and the high-frequency detail components (LH, HL, HH). This process can be expressed as:
\begin{equation}
H:y \rightarrow \{ h_{LL}(y), h_{LH}(y), h_{HL}(y), h_{HH}(y) \},  \label{eq12}
\end{equation}
where  $H$ represents the wavelet transform, $h_{LL}$  represents the low-frequency component after the transformation, $h_{LH}$, $h_{HL}$ and   $h_{HH}$   correspond to the high-frequency detail components in the vertical, horizontal and diagonal directions respectively. Let the three high-frequency components be denoted as $Y_{h}=\{ h_{LH}(y),h_{HL}(y),h_{HH}(y) \}$. Training with the dynamic features $\tilde{Y}_{h}$ obtained through random sampling of the high-frequency sub-bands in the wavelet domain effectively enhances the network's ability to represent high-frequency details. The VE-SDE is represented as follows:
\begin{equation}
d\tilde{Y}_{h}= \sqrt{d[\sigma_{h}^{2}(t)]/dt} dw_{h},   \label{eq13}
\end{equation}
where $\sigma_{h}(t)>0$  represents a monotonically increasing function and signifies the time-varying escalating scale function for noise.  The  prior learning stage for random wavelet high-frequency components involves training the high-frequency components through the neural network $\textbf{s}_{\theta_{h}}(\tilde{Y}_{h}(t),t)$. Gaussian noise is gradually introduced into the randomly selected high-frequency sub-bands in the wavelet domain to optimize the network parameters. The optimization objective function of this process is: 
\begin{equation}
\begin{split}
\theta^{\ast}_{h} = \underset{\theta_{h}}{\arg\min} \ \mathbb{E}_{t} \{ \lambda_{t} \mathbb{E}_{\tilde{Y}_{h}(0)}  \mathbb{E}_{\tilde{Y}_{h}(t)| \tilde{Y}_{h}(0)} [ \| \textbf{s}_{\theta_{h}}(\tilde{Y}_{h}(t),t)  \\
 -\nabla_{\tilde{Y}_{h}(t)} \log p_{0t}(\tilde{Y}_{h}(t) | \tilde{Y}_{h}(0) )   \| ^{2} _{2} ] \}.   
\end{split}
  \label{eq14}
\end{equation}
With sufficient data and model capacity, score matching ensures that $ \textbf{s}_{\theta_{h}}(\tilde{Y}_{h}(t),t) \approx \nabla_{\tilde{Y}_{s}(t)} \log p_{0t} $ for almost all  $\tilde{Y}_{h}$ and $t$.

To sum up,  $ \textbf{s}_{\theta_{s}}(\tilde{Y}_{s}(t),t) $  and $\textbf{s}_{\theta_{h}}(\tilde{Y}_{h}(t),t)$  can accurately learn the data distribution of real images in the projection domain and the high-frequency data distribution of sinogram wavelet. This method not only enhances the model's ability to identify the overall structure of the image but also improves its sensitivity to details such as textures. By adopting this dual-model strategy, the model is further encouraged to learn and approximate the underlying probability distribution of the data.

\subsection{Cascade Reconstruction of Sinogram and Wavelet}

In iterative reconstruction, the image generation process can essentially be formulated as the inverse problem of  SDEs. By incorporating prior knowledge from SRM and SHD, the reconstruction phase captures not only comprehensive global structures but also preserves fine-grained local details. Leveraging these priors along with data consistency constraints, the SDE framework employs a Predictor-Corrector (PC) sampler to enable stable and efficient reverse-time evolution from noise to data distribution, thereby producing high-quality images. Reverse-time SDE  can be characterized  as follows:
\begin{equation}
d\hat{y} = [f(\hat{y},t)-g^{2}(t) \nabla_{\hat{y}} \log  p_{t}(\hat{y})]dt +g(t)d \bar{w},  
\label{eq15}
\end{equation}
where $dt$ is an infinitesimal negative time step, $\bar{w}$  is a standard Brownian motion with time flows backwards from $T \rightarrow 0$. $\nabla_{\hat{y}} \log p_{t}(\hat{y})$  is the score for each marginal distribution, $\hat{y}=\{\hat{y}_{s},\hat{y}_{h} \}$ includes the sinogram and its wavelet high-frequency regions.  $\nabla_{\hat{y}} \log p_{t}(\hat{y})$  can be estimated by training a time-dependent scoring network $\textbf{s}_{\theta}(\hat{y},t)=\{\textbf{s}_{\theta}(\hat{y}_{s},t),\textbf{s}_{\theta}(\hat{y}_{h},t) \} $  to satisfy $\textbf{s}_{\theta}(\hat{y},t)\simeq \nabla_{\hat{y}} \log  p_{t}(\hat{y}) $. The score function $\theta^{\ast}$  can be estimated by network training in the learning stage of prior information. The scoring estimator $\textbf{s}_{\theta}(\hat{y}_{s,t},t)$  can replace Eq. (\ref{eq15}) to approximate the solution of the score function:
\begin{equation}
d\hat{y} = [f(\hat{y},t)-g^{2}(t) \textbf{s}_{\theta}(\hat{y},t)]dt +g(t)d \bar{w}.  
\label{eq16}
\end{equation}

During the joint iterative reconstruction stage, the introduction of regularized prior information effectively guides the model to progressively integrate the global consistency features from the projection domain with the fine-grained details embedded in the high-frequency components of the wavelet domain, thereby enabling the collaborative restoration of structural and textural information. The overall optimization objective can be formulated as follows:
\begin{equation}
\begin{split}
\{ \hat{y}_{s}^{\ast}, \hat{y}_{h}^{\ast} \}=\underset{\hat{y}_{s}, \hat{y}_{h}}{\arg\min} \ \|P(\wedge)y- \hat{y}_{s}^{\ast}\|_{2}^{2}+ \beta \| \tilde{H}_{h} [P(\wedge)y ] \\
 - \hat{y_{h}} \| ^{2}_{2} + \nu_{s}R_{s}(\hat{y}_{s}) + \nu_{h}R_{h}(\hat{y}_{h}),
\end{split}
  \label{eq17}
\end{equation}
where  $\hat{y}_{s}$ and  $\hat{y}_{h}$ represent the projection domain and the wavelet domain data in the reconstruction stage, respectively.        $\tilde{H}_{h}[\cdot]$ represents extracting the high-frequency components of the wavelet domain from the sinogram. The hyperparameters $\nu_{s}$ and $\nu_{h}$ are used to balance data consistency and regularized priors.  High-frequency information extraction is expressed as follows:
\vspace{-0.5 em}
\begin{equation}
\tilde{H}_{h}[\cdot] = \tilde{H}[\cdot] / \tilde{H}_{l}[\cdot],  \label{eq18}
\end{equation}
where  $\tilde{H}[\cdot]$ represents the full-frequency information, $\tilde{H}_{l}[\cdot]$  represents the low-frequency component and $\tilde{H}_{h}[\cdot]$  represents the high-frequency component. Through the iterative optimization mechanism of SRM and SHD, a smooth transition from global consistency to detail precision is effectively achieved.

\subsection{Sinogram Generation}

The sinogram generation stage enhances the global structural information in the image, thereby ensuring the consistency and integrity of the data during the processing. In this stage, it can be further refined into two interrelated subtasks, which respectively focus on the modeling of global information and the expression and optimization of structural features, so as to improve the expressive ability of the overall data and the reconstruction accuracy. This stage can be decomposed into the following two sub-problems:  
\vspace{-0.5 em}
\begin{flalign}
&\hat{y}_{s}^{t-\frac{1}{2}} =  \underset{\hat{y}_{s}}{\arg\min} \ \| \hat{y}_{s}-P(\wedge)y  \| ^{2}_{2} + \mu_{s}\| \hat{y}^{t}_{h} - \tilde{H}_{h}[\hat{y}_{s}]  \|_{2}^{2}, \\
&\hat{y}_{s}^{t-1} =  \underset{\hat{y}_{s}}{\arg\min} \ \|\hat{y}_{s}-\hat{y}_{s}^{t-\frac{1}{2}}  \| ^{2}_{2} + \nu_{s}R_{s}(\hat{y}_{s}).
\label{eq19-20}
\vspace{-0.5 em}
\end{flalign}

On the basis of ensuring the consistency between the generated data and the original observed data, the data consistency constraint term enhances the stability of the model optimization process. 
\vspace{-0.5 em}
\begin{equation}
\hat{y}_{s} = (1-P(\wedge))\hat{y}_{s} + P(\wedge)y .  
\label{eq21}
\vspace{-0.5 em}
\end{equation}

The predictor serving as a solver for the Variance Exploding type of Stochastic Differential Equation, generates estimates for updates in each reconstruction iteration and numerically solves this SDE through the reverse diffusion process. In this process, the pre-trained model $\textbf{s}_{\theta_{s}}$ is utilized to sample the reverse SDE, thereby achieving the gradual generation of samples. The discretization process can be expressed as follows:
\begin{equation}
\hat{y}_{s}^{t-1} = \hat{y}_{s}^{t-\frac{1}{2}} + (\delta_{t}^{2}-\delta_{t-1}^{2})\textbf{s}_{\theta_{s}}(\hat{y}_{s}^{t-\frac{1}{2}},t) + \sqrt{\delta_{t}^{2}-\delta_{t-1}^{2}}z   ,  \label{eq22}
\end{equation}
where $\delta_{t}$  represents a monotonically increasing function with respect to time $t$. $z\sim \mathcal{N}(0,1)$  is a gaussian distribution following random noise. The corrector uses Langevin dynamics to convert the initial sample $\hat{y}_{s}(0)$ to the final sample $\hat{y}_{s}(t)$, the steps to implement the ``corrector'' are as follows:
\begin{equation}
\hat{y}_{s}^{t-1} = \hat{y}_{s}^{t-1} + \varepsilon_{t-1}s_{\theta_{s}}(\hat{y}_{s}^{t-1},t)+ \sqrt{2\varepsilon_{t-1}}z  ,  \label{eq23}
\end{equation}
where $\varepsilon > 0$  is the step size and the above equation is repeated for $t=T-1,\cdots,0$. The solution is optimized by alternating iterations of the ``predictor'' and ``corrector'' steps above to achieve convergence.

\subsection{High-frequency generation in wavelet domain}

After the preliminary reconstruction is completed in the projection domain during the first stage, the obtained sinogram is mapped to the wavelet domain. In this domain, the image is decomposed at multiple scales, from which the high-frequency components in three directions are extracted, and the low-frequency components are retained simultaneously, so as to fully capture the detailed structures and the overall contours in the image. The process can be described as follows: 
\begin{equation}
\begin{split}
\hat{y}_{h}^{t-\frac{1}{2}} & = \tilde{H}[\hat{y}_{s}^{t-1}] / \tilde{H}_{l}[\hat{y}_{s}^{t-1}] =  \tilde{H}_{h}[\hat{y}_{s}^{t-1}]   \\
 &= \{ \hat{y}_{LH}^{t-\frac{1}{2}} , \hat{y}_{HL}^{t-\frac{1}{2}} , \hat{y}_{HH}^{t-\frac{1}{2}}   \}.
\end{split}
  \label{eq24}
\end{equation}

The data consistency of the high-frequency components in the wavelet domain to ensure the reliability of the data:
\begin{equation}
\hat{y}_{h,i}= \tilde{H}_{h}[(1-P(\wedge))\hat{y}_{s} + P(\wedge)y] .  \label{eq25}
\end{equation}

The  ``predictor-corrector'' framework in the VE-SDE is also applied to the reconstruction process of each high-frequency channel. Specifically, in the ``predictor'' step, the  network $\textbf{s}_{\theta_{h}}$ is utilized to reconstruct each high-frequency component, and the process can be described as follows:
\begin{equation}
\hat{y}_{h,i}^{t-1}  = \hat{y}_{h,i}^{t-\frac{1}{2}} + ( \delta_{t}^{2}-\delta_{t-1}^{2})\textbf{s}_{\theta_{h}}( \hat{y}_{h,i}^{t-\frac{1}{2}},t) + \sqrt{\delta_{t}^{2} - \delta_{t-1}^{2}}z .  \label{eq26}
\end{equation}

Then, perform the ``corrector'' step as well, as follows:
\begin{equation}
\hat{y}_{h,i}^{t-1}  = \hat{y}_{h,i}^{t-1} + \varepsilon_{t-1}\textbf{s}_{\theta_{h}}(\hat{y}_{h}^{t-1},t) + \sqrt{2\varepsilon_{t-1}}z .  \label{eq27}
\end{equation}

\subsection{Domain transform stage}

Finally, this process integrates the previously retained low-frequency information with the optimized high-frequency components. By fusing information across different frequency levels, a more comprehensive image representation is achieved. Subsequently, the combined data is mapped back to the projection domain through the inverse wavelet transform, as detailed below:
\begin{equation}
\hat{y}= \tilde{H}^{T}[\tilde{H}_{l}[\hat{y}_{s}], \tilde{H}_{h}[\hat{y}_{s}]]. 
 \label{eq28}
\end{equation}

Upon completion of the iteration, the final reconstructed image is generated by applying the Filtered Back Projection (FBP) algorithm for back projection. The  expression for this process is as follows:
\begin{equation}
 \tilde{x} = FBP(\hat{y}). 
 \label{eq29}
\end{equation}

In summary, the iterative reconstruction stage of SWARM is shown in Fig. \ref{train+test}(c).  In the actual reconstruction, through the iterative updates of the numerical SDE solver and the Langevin dynamics, high-quality full-view projection data is gradually obtained.  Algorithm \ref{alg1} provides a detailed description of both the training process and the reconstruction stage.

\begin{algorithm}[H]
\caption{Training and Iterative Reconstruction Process.}
\begin{algorithmic}
\STATE \hspace{-0.4cm} \textbf{Dataset:} Load image data, generate projection data via FP forward projection.
\STATE \hspace{-0.4cm} \textbf{SRM Training:} Sample a projected data $y$, apply a random mask $\hat{m}$ to generate $\tilde{Y}_s = y \odot \hat{m}$, and train the SRM network $\textbf{s}_{\theta_s}(\tilde{Y}_s, t)$.
\STATE \hspace{-0.4cm} \textbf{SHD Training:} Sample a projected data $y$, randomly select one high-frequency wavelet component$ \tilde{Y}_{h} \leftarrow \{ h_{LH}(y), h_{HL}(y), h_{HH}(y) \} $, and train the SHD network $\textbf{s}_{\theta_h}(\tilde{Y}_h, t)$.
\STATE \hspace{-0.4cm} \textbf{Iterative Reconstruction Process:}
\STATE \hspace{-0.4cm}  \textbf{Setting:} $\textbf{s}_{\theta_{s}}$, $\textbf{s}_{\theta_{h}}$, $\sigma$, $\varepsilon$;
\STATE \hspace{-0.4cm}  1: \ For $t=T-1$ to $0$ do:
\STATE \hspace{-0.4cm}  2:  \hspace{0.3cm}  $\hat{Y}_{s}^{t-\frac{1}{2}} \leftarrow predictor(\hat{Y}_{s}^{t}, \sigma_{t-1}, \sigma_{t}, \textbf{s}_{\theta_{s}}) $;
\STATE \hspace{-0.4cm}  3: \hspace{0.3cm} Update $\hat{Y}_{s}^{t-\frac{1}{2}}$ with Eq. (\ref{eq21});
\STATE \hspace{-0.4cm}  4: \hspace{0.3cm}  $\hat{Y}_{s}^{t-1} \leftarrow corrector(\hat{Y}_{s}^{t-\frac{1}{2}}, \sigma_{t-1}, \varepsilon_{t-1}, \textbf{s}_{\theta_{s}}) $;
\STATE \hspace{-0.4cm}  5: \hspace{0.3cm}  Update $ \hat{Y}_{s}^{t-1} $ with Eq. (\ref{eq21});
\STATE \hspace{-0.4cm}  6: \hspace{0.3cm} For each subband $i \in \{LH, HL, HH\}$ do:
\STATE \hspace{-0.4cm}  7: \hspace{0.6cm} $\hat{Y}_{h,i}^{t-\frac{1}{2}} \leftarrow predictor(\hat{Y}_{h,i}^{t}, \sigma_{t-1}, \sigma_{t}, \textbf{s}_{\theta_{h}})$;
\STATE \hspace{-0.4cm}  8: \hspace{0.6cm} Update $\hat{Y}_{h,i}^{t-\frac{1}{2}}$ via Eq. (\ref{eq25});
\STATE \hspace{-0.4cm}  9: \hspace{0.5cm} $\hat{Y}_{h,i}^{t-1} \leftarrow corrector(\hat{Y}_{h,i}^{t-\frac{1}{2}}, \sigma_{t-1}, \varepsilon_{t-1}, \textbf{s}_{\theta_{h}})$;
\STATE \hspace{-0.4cm}  10: \hspace{0.5cm} Update $\hat{Y}_{h,i}^{t-1}$ via Eq. (\ref{eq25});
\STATE \hspace{-0.4cm}  11: \hspace{0.2cm} End for
\STATE \hspace{-0.4cm}  12: \hspace{-0.1cm} End for
\STATE \hspace{-0.4cm}  13: \hspace{-0.1cm} \textbf{return} $\tilde{x}$.
\end{algorithmic}
\label{alg1}
\end{algorithm}

\section{Experiment}

\subsection{Data Specification}

\subsubsection{AAPM Challenge Data}

The dataset used for training and testing is part of the AAPM Low-Dose Challenge\cite{AAPM} provided by the Mayo Clinic.  In this study,  9 patients were used for   The distance from the center of rotation to the source and detector is set at 40 cm and 40 cm, respectively. The detector is 41.3 cm wide, has 720 detector elements, and a total of 720 projection views are evenly distributed.

\subsubsection{CIRS Phantom Data}

We use a dataset of high-quality CT scans to further analyze the performance of the proposed method, each with dimensions of 512×512×100 voxels, and a voxel dimension of 0.78×0.78×0.625 mm$^{3}$. The data set was collected using the GE Discovery HD750 CT scanner combined with a bionic model provided by CIRS. The vacuum tube current is set to 600 mAs, the source-wheelbase separation of the CT system is 573mm, and the source-detector distance is 1010 mm. 

\subsubsection{Dental Arch Data}

The data set is the clinical data collected by JIROX Dental CBCT device produced by  YOFO (Hefei) Medical Technology Co., Ltd.  The source-to-image distance (SID) of the device is 1700 mm, and the source-to-detector distance (SAD) is 1500 mm.  The device operates at a voltage of 100 kV and a current of 6 mA. The detector array consisted of 768 × 768 elements, with each detector element measuring 0.2 × 0.2 mm. The dataset consists of 20 cases, and each case provides 200 slices.  The dataset consists of full-view sinogram and their corresponding FBP reconstructions, with 512 × 512 image matrix. We randomly selected 1 case from the obtained image matrices.

\subsection{Implementation Details}

In our experiments, the model is trained using the Adam optimizer with a learning rate of $10^{-3}$ and Kaiming weight initialization. Model hyperparameters are set as follows: VE-SDE parameters $\sigma_{\min}=0.01$ and $\sigma_{\max}=378$. 
The noise step schedule $\epsilon_t$ follows the VE-SDE~\cite{song2020score}, and the predictor–corrector (PC) step ratio is set to 1.

Sparse-view projections were simulated using inverse masks at 10, 20, 30, 60, 90, 120, and 180 views. The masks were generated under the sparse scanning model by uniformly random sampling without fixing random seeds, and independently applied to each sample. The 2D discrete wavelet transform (DWT) uses the Haar wavelet with two decomposition levels. Boundary handling is done via symmetric extension with matrix construction and cropping, ensuring stability and boundary preservation. High-frequency components in each sub-band are randomly selected with uniform probability.

In the fan-beam CT experiments, ray-driven algorithms \cite{article1, article} simulated projection data, using ODL \cite{rajpurkar2022ai}. The projection geometry includes 720 projection angles over $[0, 2\pi]$ and 720 detector elements over $[-360, 360]$~mm. Both the SOD and the SID are 500~mm, and the reconstruction domain is discretized on a $512 \times 512$ grid. The reference images are generated from 720 projection views using the FBP algorithm. Our source code is publicly available on GitHub and can be accessed via the following link:  https://github.com/yqx7150/SWARM. To evaluate the proposed method with different sparse views, CT images were reconstructed using 60, 90, and 120 projections. Performance was quantitatively assessed using PSNR, SSIM, and MSE.

\begin{table*}[h]
\centering
\caption{Reconstruction PSNR/SSIM/MSE$(10^{-3})$ of AAPM Challenge Data Using Different Methods at   60, 90, 120 Views.}
 \vspace{-5pt}
\begin{tabular}{ccccccc}
\toprule
Views & FBP\cite{brenner2007computed} & FBPConvNet\cite{McCann2017Deep} & HDNet\cite{Hu2021Hybrid-Domain} &  GMSD\cite{Guan2023Generative} & SWORD\cite{10403850} & SWARM \\
\midrule
60 & 23.28/0.5957/4.815 & 34.23/0.9564/0.402 & 35.28/0.9706/0.345 & 36.29/0.9684/0.276 & 37.09/0.9738/0.213 & \textbf{38.43}/\textbf{0.9809}/\textbf{0.188} \\
90 & 26.32/0.7088/2.388 & 36.25/0.9611/0.260 & 38.45/0.9809/0.164 &  39.22/0.9800/0.127 & 40.14/0.9835/0.103 & \textbf{41.91}/\textbf{0.9882}/\textbf{0.067} \\
120 & 28.50/0.7961/1.447 & 38.03/0.9692/0.161 & 39.21/0.9851/0.151  & 40.53/0.9846/0.097 & 42.04/0.9884/0.066 & \textbf{42.80}/\textbf{0.9904}/\textbf{0.058} \\
\bottomrule
\end{tabular}
\label{AAPM_IMG_TABLE}
\end{table*}


\begin{figure*}[h]
     \centerline{\includegraphics[width=0.9\textwidth]{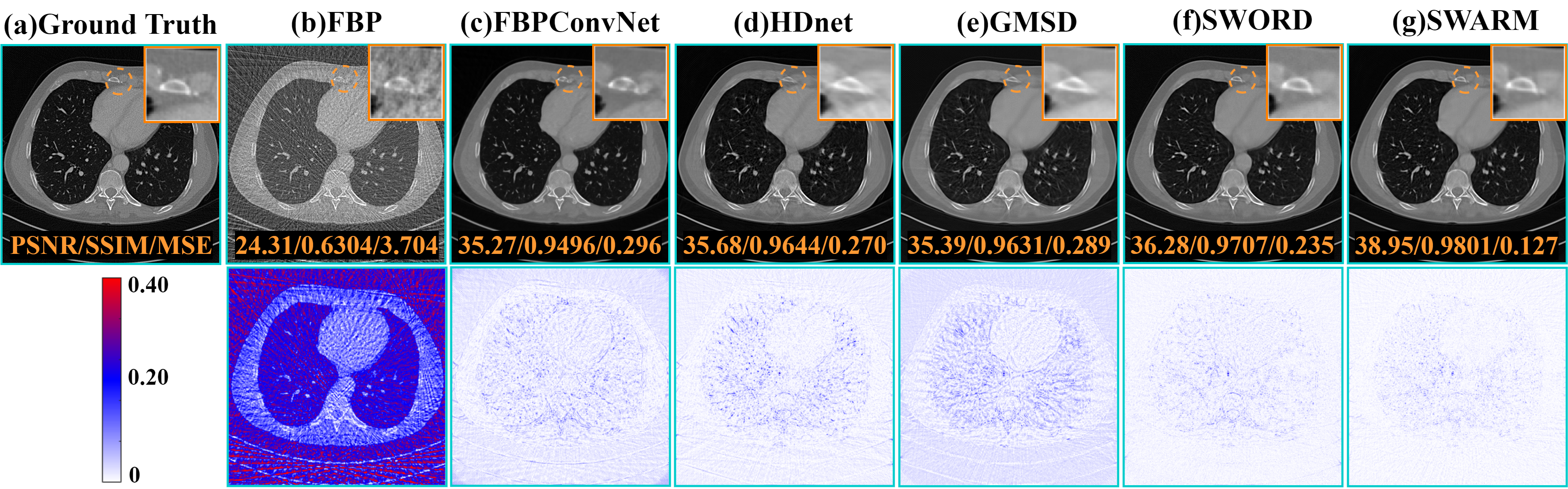}} 
    \caption{Reconstruction images from 90 views using different methods with AAPM challenge data. (a) The reference image versus the images reconstructed by (b) FBP, (c) FBPConvNet, (d) HDNet, (e) GMSD, (f) SWORD, and (g) SWARM. The display window is [-480, 945] HU. The second line is the residual between the reference image and the reconstructed image.}
    \label{aapm_90}
\end{figure*}

\begin{table*}[h]
\centering
\caption{Reconstruction PSNR/SSIM/MSE$(10^{-3})$ of CIRS Phantom Data Using Different Methods at 60, 90, 120 Views.}
 \vspace{-5pt}
\begin{tabular}{ccccccc}
\toprule
Views & FBP\cite{brenner2007computed} & FBPConvNet\cite{McCann2017Deep} & HDNet\cite{Hu2021Hybrid-Domain} &  GMSD\cite{Guan2023Generative} & SWORD\cite{10403850} & SWARM \\
\midrule
60 & 17.15/0.5057/19.294 & 24.25/0.9308/3.762 & 32.53/0.9750/0.562  & 33.07/0.9744/0.499 & 32.89/0.9750/0.520 & \textbf{38.38}/\textbf{0.9887}/\textbf{0.146} \\
90 & 21.78/0.6215/6.636 & 29.60/0.9320/1.102 & 37.21/0.9881/0.193  & 37.75/0.9866/0.170  & 39.21/0.9901/0.124 & \textbf{44.73}/\textbf{0.9958}/\textbf{0.034} \\
120 & 25.08/0.6940/3.102 & 31.19/0.9444/0.764 & 40.20/0.9916/0.970  & 39.86/0.9915/0.104 & 41.30/0.9935/0.074 & \textbf{46.80}/\textbf{0.9969}/\textbf{0.021} \\
\bottomrule
\end{tabular}
\label{CIRS_IMG_TABLE}
\end{table*}

\begin{figure*}[htbp]
\centering
\centerline{\includegraphics[width=0.9\textwidth]{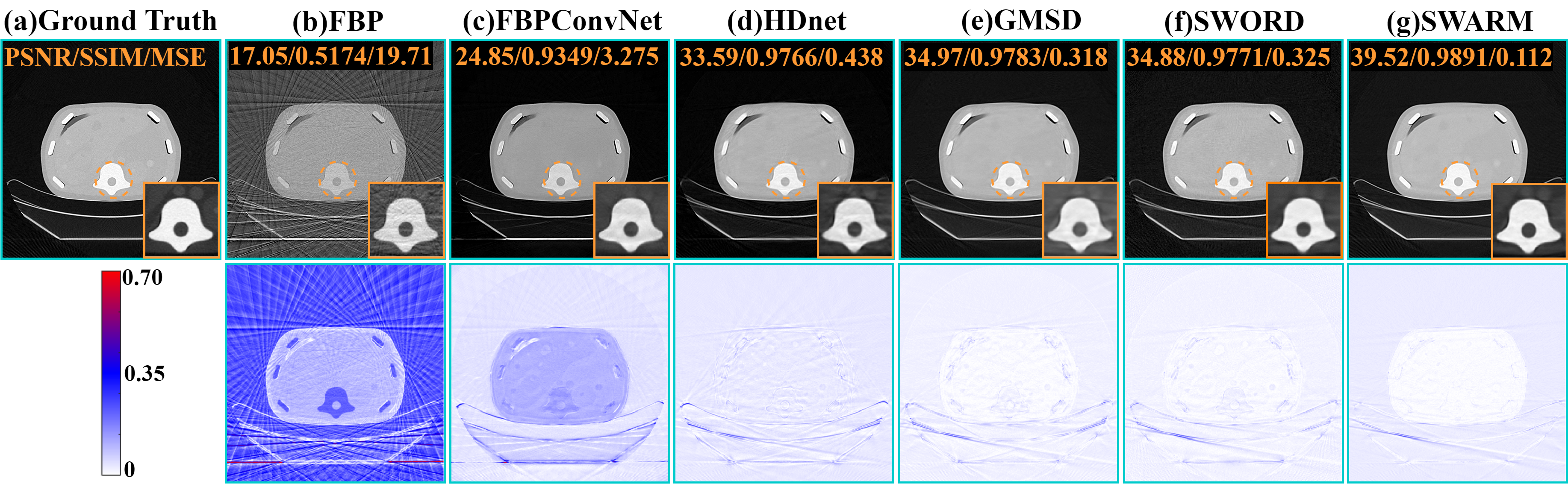}} 
\caption{Reconstruction images from 60 views using different methods with CIRS phantom data. (a) The reference image versus the images reconstructed by (b) FBP, (c) FBPConvNet, (d) HDNet, (e) GMSD, (f) SWORD, and (g) SWARM. The display window is [675, 1300] HU. The second line is the residual between the reference image and the reconstructed image.}
    \label{cirs_60}
\end{figure*}

\begin{table*}[h]
\centering
\caption{Reconstruction PSNR/SSIM/MSE$(10^{-3})$ of Dental Arch Data Using Different Methods at 60, 90, 120 Views.}
\vspace{-5pt}
\begin{tabular}{ccccccc}
\toprule
Views & FBP\cite{brenner2007computed} & FBPConvNet\cite{McCann2017Deep} & HDNet\cite{Hu2021Hybrid-Domain} & GMSD\cite{Guan2023Generative} & SWORD\cite{10403850} & SWARM \\
\midrule
60 & 25.55/0.7015/2.818 & 31.86/0.9492/0.678 & 31.20/0.9658/0.972  & 33.45/0.9678/0.483 & 36.33/0.9757/0.251 & \textbf{40.22}/\textbf{0.9911}/\textbf{0.099} \\
90 & 28.12/0.7917/1.574 & 34.04/0.9574/0.406 & 36.55/0.9232/0.248  & 37.70/0.9826/0.183 & 39.93/0.9890/0.109 & \textbf{44.63}/\textbf{0.9955}/\textbf{0.037} \\
120 & 29.87/0.8519/1.056 & 35.55/0.9716/0.307 & 40.25/0.9904/0.102 & 40.68/0.9902/0.091 & 43.58/0.9944/0.047 & \textbf{47.11}/\textbf{0.9971}/\textbf{0.022} \\
\bottomrule
\end{tabular}
\label{Panoramic_TABLE}
\end{table*}

\begin{figure*}[htbp]
 \centering
  \centerline{\includegraphics[width=0.9\textwidth]{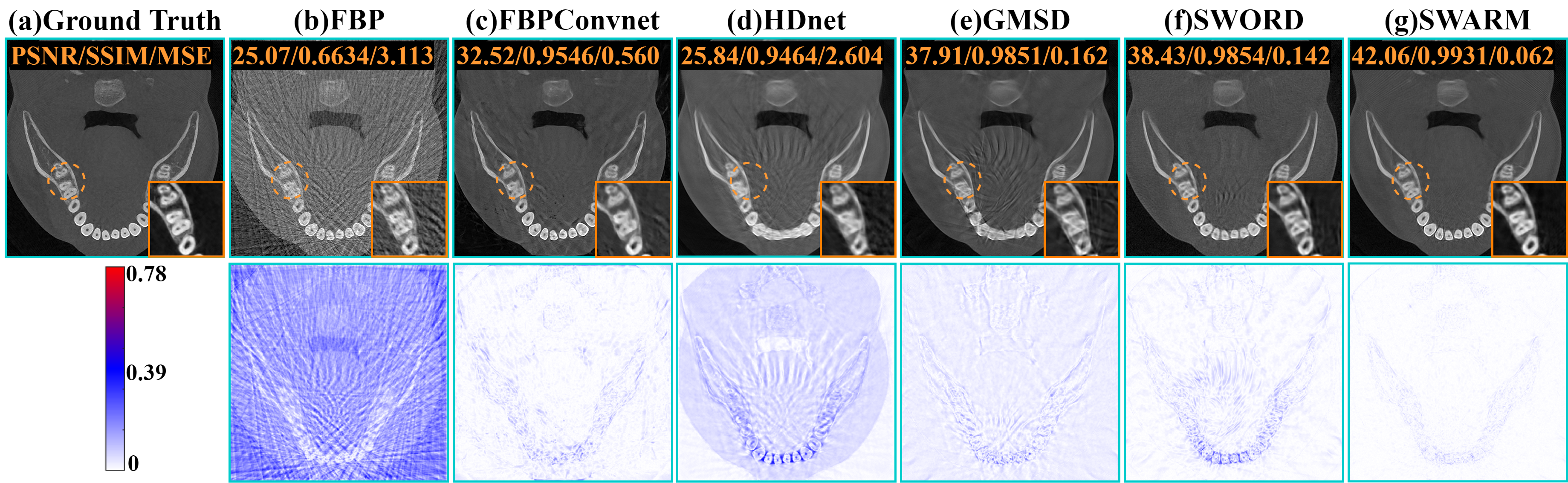}} 
    \caption{Reconstruction images from 60 views using different methods with Dental Arch data. (a) The reference image versus the images reconstructed by (b) FBP, (c) FBPConvNet, (d) HDNet, (e) GMSD, (f) SWORD, and (g) SWARM. The display window is [-60, 1300] HU. The second line is the residual between the reference image and the reconstructed image.}
    \label{yofo_60}
\vspace{-5pt}
\end{figure*}

\begin{figure*}[htbp]
    \centerline{\includegraphics[width=0.9\textwidth]{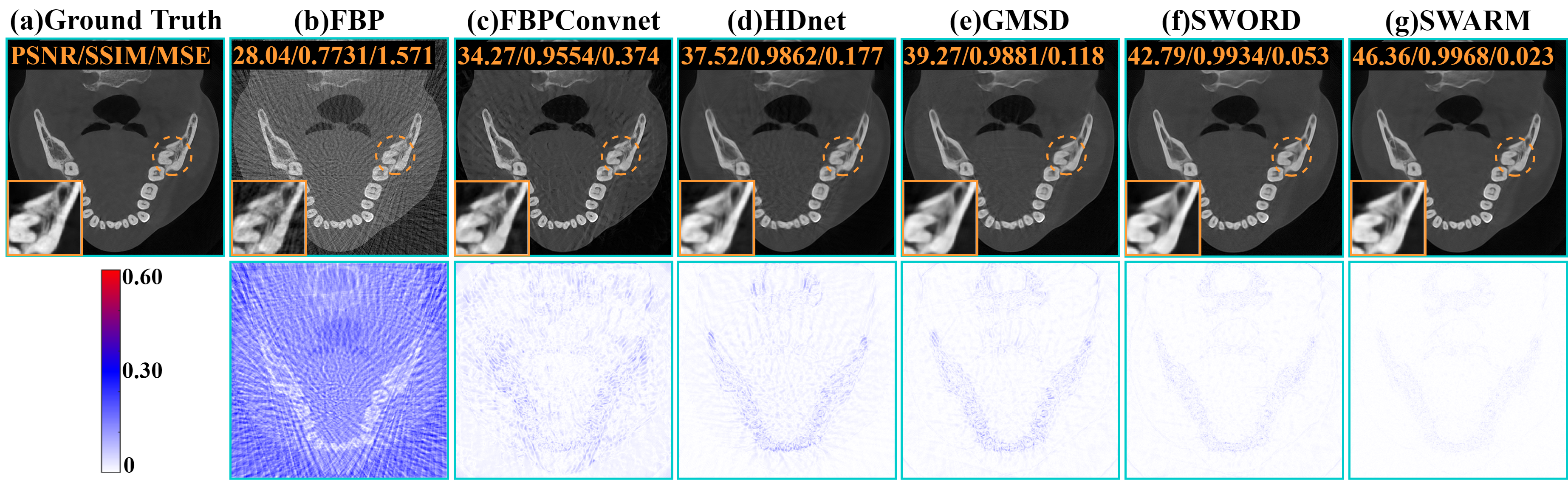}} 
    \caption{Reconstruction images from 90 views using different methods with Dental Arch data. (a) The reference image versus the images reconstructed by (b) FBP, (c) FBPConvNet, (d) HDNet, (e) GMSD, (f) SWORD, and (g) SWARM. The display window is [-500, 970] HU. The second line is the residual between the reference image and the reconstructed image.}
    \label{yofo_90}
\end{figure*}

\subsection{Reconstruction Experiments}

\subsubsection{AAPM Reconstruction Results}

In order to assess the effectiveness of the proposed algorithm, we evaluated the performance and compared some representative methods including FBP\cite{brenner2007computed}, FBPConvNet\cite{McCann2017Deep}, HDNet\cite{Hu2021Hybrid-Domain}, GMSD\cite{Guan2023Generative}, SWORD\cite{10403850}.  Sparse-view CT reconstructions were conducted using  60, 90, and 120 projection, respectively. Table \ref{AAPM_IMG_TABLE} presents the PSNR, SSIM and MSE values for the reconstruction results of the AAPM challenge dataset. The best values for reconstructed images with different projection views are highlighted in bold. SWARM  indicating its superior performance in overall image quality and signal fidelity.  The higher PSNR demonstrates that SWARM can more accurately restore the overall signal intensity of the reconstructed images, maintaining a high consistency with the ground truth. The optimal SSIM further reflects its enhanced capability in preserving structural information and local texture details. The substantial reduction in MSE further confirms the minimal reconstruction error, bringing the reconstructed images closer to the ground truth.

Fig. \ref{aapm_90} shows the visual reconstruction results of the AAPM test dataset.  FBP  exhibits noticeable edge blurring and streak artifacts, with severe loss of fine structures. Although FBPConvNet alleviates some of these artifacts, edge sharpness remains insufficient, and certain local details are still not recovered. HDNet achieves a generally smoother reconstruction, but excessive smoothing leads to blurred key structural boundaries. Similarly, GMSD and SWORD suffer from texture loss and weakened detail preservation during iterative reconstruction. In contrast, SWARM more clearly restores the edges and structural details of lung tissue, with the reconstructed images showing higher consistency with the ground truth in terms of texture, morphology, and intensity distribution. In summary, SWARM not only preserves the structural integrity of the images but also enhances the overall signal accuracy, providing a more reliable imaging foundation for clinical applications.

\subsubsection{CIRS Phantom Reconstruction Results}

To further evaluate the  proposed  method, prior knowledge was learned from the AAPM challenge dataset, and the model performance was assessed using CIRS phantom data. Table \ref{CIRS_IMG_TABLE} shows that SWARM significantly outperforms the other approaches in terms of quantitative metrics. The evaluation results of FBP and FBPConvNet on the CIRS phantom dataset are notably inferior, indicating considerable loss of image information. Although the quantitative metrics of HDNet and GMSD remain relatively stable, their reconstruction performance still shows a noticeable decline. SWORD shows relatively stable reconstruction performance in various indicators. However, it can still be observed that its reconstruction quality decreases to a certain extent in some test cases, revealing the limitation of insufficient generalization ability. In contrast, SWARM achieved the best reconstruction results, fully demonstrating its superior overall performance.

As shown in Fig. \ref{cirs_60},  FBP  suffer from prominent streak artifacts and significant loss of structural details. FBPConvNet  remains limited in restoring edge sharpness and texture details, resulting in blurred structural contours. Despite leveraging iterative optimization of dual-domain information, HDNet remains inadequate in accurately delineating internal structural boundaries. Similarly, GMSD and SWORD tend to over-smooth the reconstructed images, leading to the loss of texture information and blurring of fine structures. In contrast, SWARM demonstrates superior performance by effectively suppressing artifacts while preserving rich texture details and structural features. This results in a substantial enhancement of visual reconstruction quality, fully showcasing the method's excellent generalization capability and robustness across diverse reconstruction scenarios.

\subsubsection{Dental Arch Reconstruction Results}

To evaluate the practicality of the proposed method, the model learned prior knowledge from the abdominal dataset of the AAPM challenge and assessed the model performance using the Dental Arch data. Table \ref{Panoramic_TABLE} shows the significant advantages of  SWARM  in terms of quantitative metrics. Although other comparison methods remain relatively stable in the evaluation results or show slight improvements, they still fail to fully demonstrate their reconstruction capabilities. In contrast,  SWARM shows obvious advantages with strong generalization ability and robustness, which provides an effective idea for solving the problem of SVCT reconstruction, and is expected to provide a potential reference for clinical diagnosis in stomatology in theory.

As shown in Fig. \ref{yofo_60} and Fig. \ref{yofo_90}, SWARM demonstrates excellent structural fidelity in oral image reconstruction. In contrast, FBP, FBPConvNet, HDNet, and GMSD exhibit noticeable streak artifacts and loss of detail in the dental region, particularly around the mandibular molars, resulting in unclear tooth morphology and indistinct boundaries of periodontal tissues, which may affect clinical assessment of tooth structures and lesions. Although SWORD partially alleviates these artifacts, it still produces blurring in regions with complex dental structures, making fine features such as interproximal spaces difficult to fully resolve. By comparison, SWARM effectively suppresses streak artifacts while preserving clear boundaries and rich textural details of teeth and surrounding tissues, with particularly outstanding performance in reconstructing the mandibular molar structures. The images reconstructed by this method provide clear visualization of dental regions, offering intuitive and discernible information for clinical dentistry, thereby enhancing diagnostic accuracy and reliability in treatment planning.


\begin{figure}[!t]
\centerline{\includegraphics[width=0.95\columnwidth]{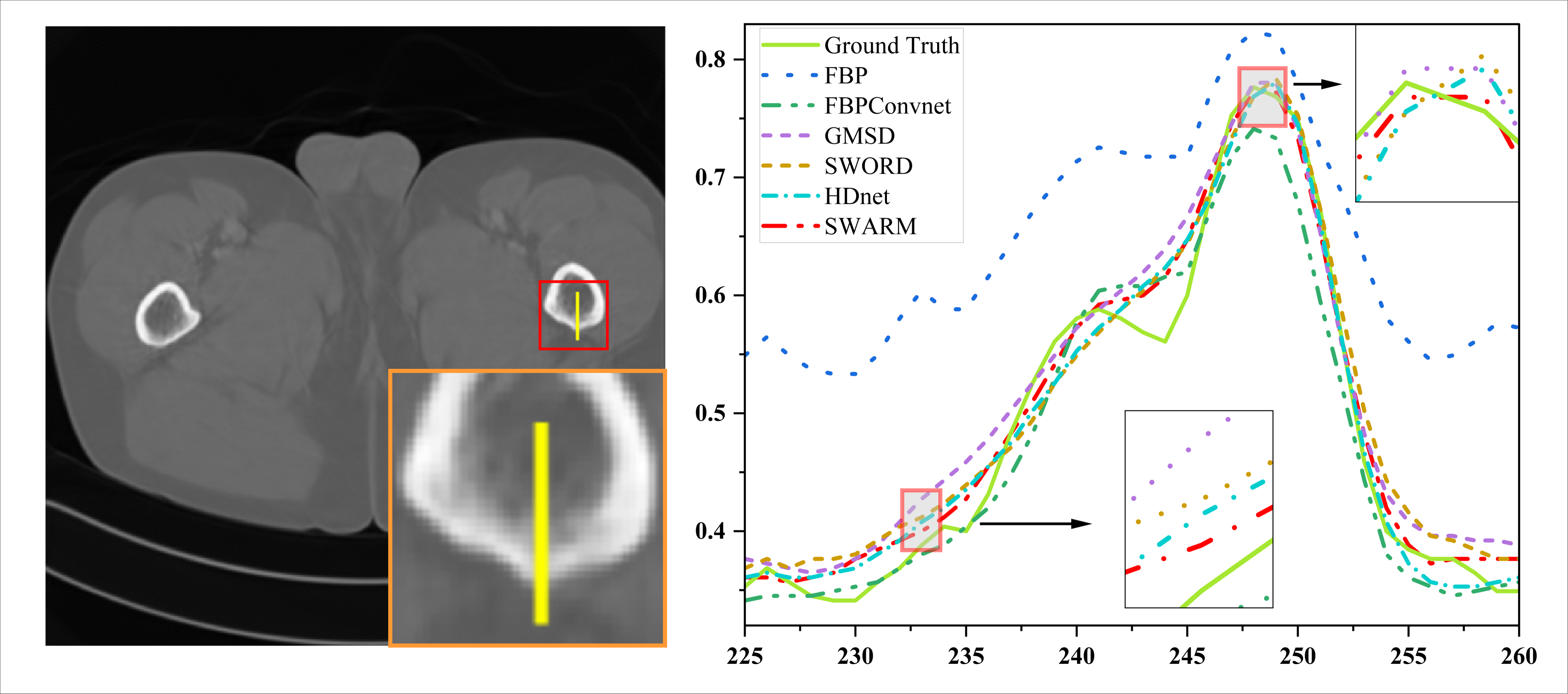}}
\caption{The intensity profiles of different methods along the specified yellow line in an example reconstructed image.}
\label{profile}
\end{figure}

\subsection{Profile Lines Analysis}

In this study, we further evaluated the reconstruction accuracy of different methods through profile analysis. This analysis quantitatively assesses the reconstruction quality by comparing the signal intensity distributions along specific cross-sectional regions between the reconstructed images and the ground truth. As shown in Fig. \ref{profile}, the signal intensity profiles of FBP, FBPConvNet, HDNet, and SWORD exhibit noticeable deviations from the ground truth, particularly in regions with sharp intensity transitions, indicating their limitations in accurately restoring structural boundaries. In contrast, GMSD and SWARM demonstrate much higher consistency with the ground truth, reflecting their superior capability in preserving local structures and intensity variations. Notably, SWARM achieves the closest numerical correspondence to the ground truth in both smooth and high-contrast regions, suggesting that it not only maintains structural integrity but also delivers superior quantitative accuracy—an attribute that is particularly crucial for medical diagnosis. These results further validate the robustness and reliability of the proposed SWARM method in reconstructing clinically relevant image details.

\begin{figure*}[h]
\centering
  \centerline{\includegraphics[width=0.95\textwidth]{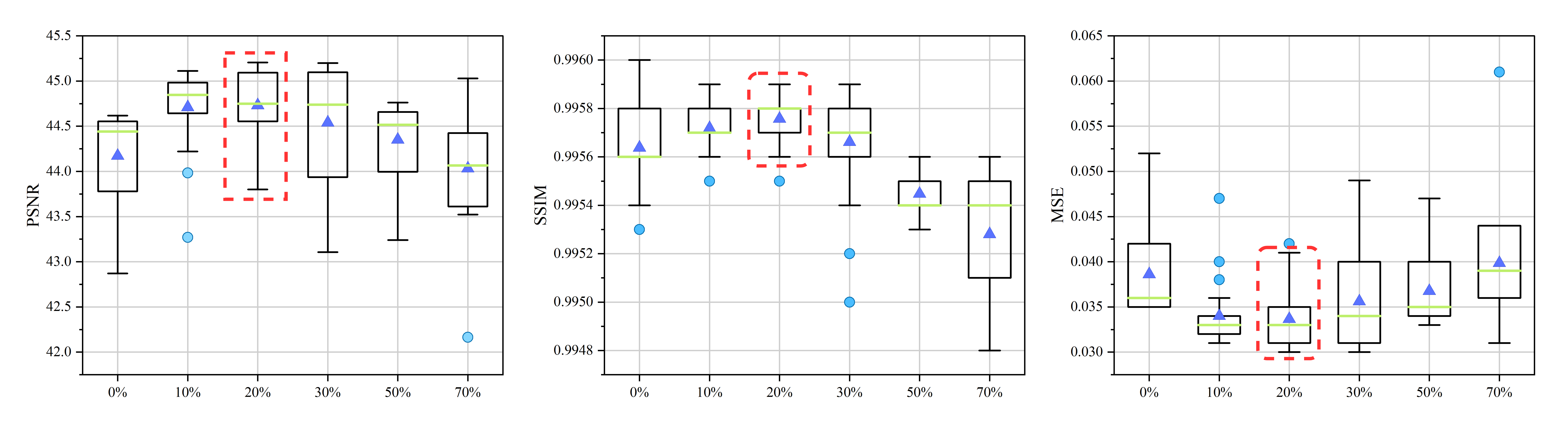}} 
    \caption{Boxplot of Performance on CIRS Test Dataset Under Different Mask Rates (0\%, 10\%, 20\%, 30\%, 50\%, 70\%).}
    \label{maskratio}
\end{figure*}


\subsection{Ablation and Generalization Study}

To evaluate the effectiveness of each component, we performed ablation experiments on each method. The comprehensiveness and accuracy of the assessment is ensured by testing the impact of each component separately.

\subsubsection{Mask Ratio Analysis}

To systematically evaluate the impact of different mask sparsity ratios on image reconstruction performance, ablation experiments were conducted with mask ratios of 0\%, 10\%, 20\%, 30\%, 50\%, and 70\%, as illustrated in Fig. \ref{maskratio}. As the masking ratio increases, both PSNR and SSIM initially improve, plateau, and then decline, while MSE demonstrates a typical “U-shaped” curve-decreasing initially and rising again beyond a certain point. This trend suggests that extremely low or high masking ratios hinder the model’s ability to effectively infer missing information, resulting in suboptimal reconstruction quality. In contrast, a moderate masking level (e.g., 20\%) introduces a beneficial degree of sparsity that stimulates the model’s prior learning capacity, achieving an optimal balance between reconstruction quality and robustness.
These findings further validate the hypothesis proposed in the methodological motivation: introducing an appropriate level of random masking enhances the model’s ability to complete unobserved regions, thereby improving overall reconstruction performance. This provides clear experimental evidence and guidance for selecting optimal mask ratios in future applications.

\subsubsection{Different Components in SWARM}

\begin{figure}[!t]
\centerline{\includegraphics[width=\columnwidth]{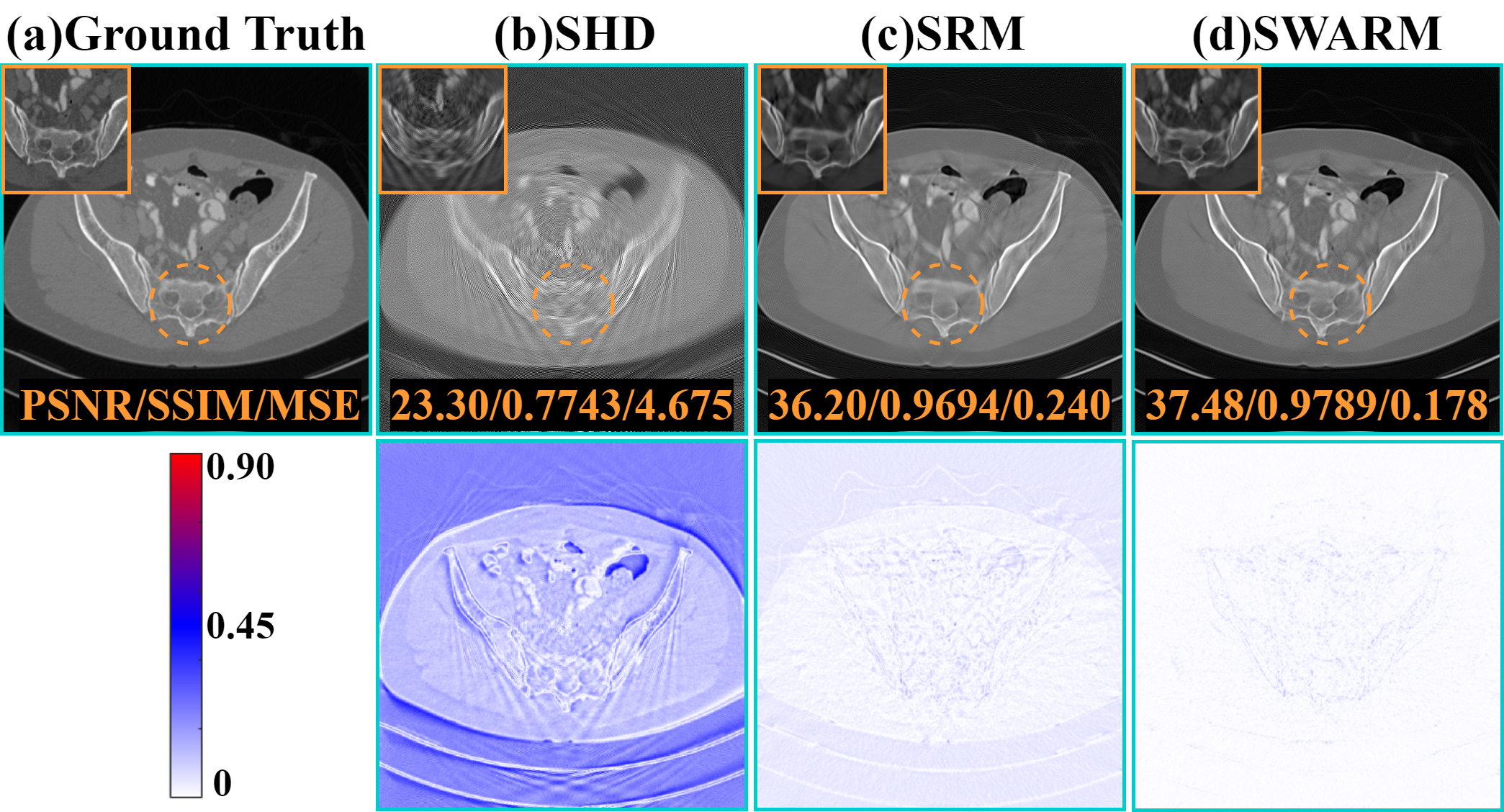}}
\caption{Reconstruction images obtained from 60 views using different methods. (a) The reference image versus the images reconstructed by (b) SHD, (c) SRM, (d) SWARM. The display window is [-180, 1300] HU. The second line is the residual between the reference image and the reconstructed image.}
\label{ablation_pic}
\end{figure}

SWARM employs a dual-diffusion model for iterative reconstruction. During this process, SRM is trained using a random mask strategy, while SHD is trained based on random high-frequency components from the sinogram. The quantitative analysis in Table \ref{ablation_table} demonstrates that SWARM significantly improves reconstruction quality. Fig. \ref{ablation_pic} further illustrates the impact of the model combination. Qualitative results show that the cascade of SRM and SHD in iterative reconstruction effectively leverages their complementary advantages, leading to a significant enhancement in image reconstruction quality. This outcome validates the effectiveness of our approach, showing that combining the strengths of both models achieves superior reconstruction performance.

\subsubsection{Effectiveness of Random Masking and High-Frequency of Sinogram}

\begin{figure}[!t]
\centerline{\includegraphics[width=\columnwidth]{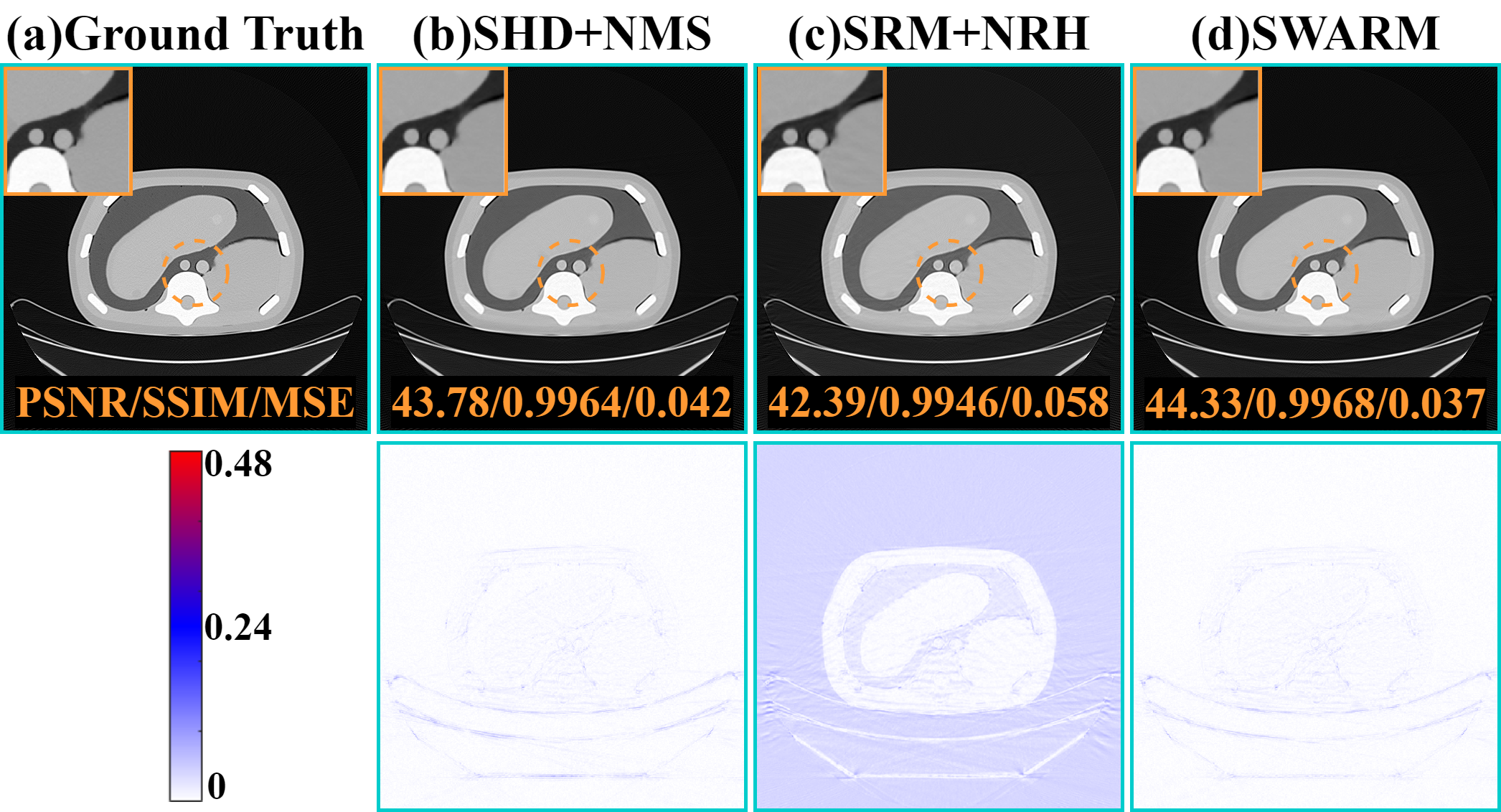}}
\caption{Reconstruction images obtained from 90 views using different methods. (a) The reference image versus the images reconstructed by (b) SHD+NMS, (c) SRM+NRH, (d) SWARM. The display window is [-180, 1370] HU. The second line is the residual between the reference image and the reconstructed image.}
\label{generalization_pic}
\end{figure}

As shown in Table \ref{generalization_table} and Fig. \ref{generalization_pic}, we studied the effects of the random masking strategy and randomness based on high-frequency components in the wavelet domain from both quantitative metrics and visual perspectives. NMS represents the non-masked condition for the sinogram, while NRH represents the non-random wavelet condition for the sinogram. Generalization experiments conducted on the CIRS phantom data further confirmed that the incorporation of random masking not only significantly improved the model's generalization ability but also enhanced its robustness, thereby validating the effectiveness of the proposed method. Additionally, the strategy of randomly selecting high-frequency components as network training inputs demonstrated good stability, effectively boosting the model's generalization capability and improving overall reconstruction efficiency.

\begin{table}[htbp]
    \centering
    \caption{Ablation Study of Two-Stage Model on AAPM Challenge Data.}
    \vspace{-5pt}
    \begin{tabular}{c|c|ccc}
        \toprule
        Methods & Views & PSNR(dB) & SSIM & MSE($10^{-3}$) \\
        \midrule
        \multirow{3}{*}{SHD} 
        & 60 & 24.10 & 0.7866 & 3.993 \\
        & 90 & 26.02 & 0.8451 & 2.567 \\
        & 120 & 27.77 & 0.8885 & 1.715 \\
        \cmidrule{1-5} 
        \multirow{3}{*}{SRM} 
        & 60 & 36.44 & 0.9691 & 0.253 \\
        & 90 & 38.77 & 0.9772 & 0.151 \\
        & 120 & 40.44 & 0.9830 & 0.100 \\
        \cmidrule{1-5} 
        \multirow{3}{*}{SWARM} 
        & 60 & \textbf{38.43} & \textbf{0.9809} & \textbf{0.188} \\
        & 90 &  \textbf{41.91} &  \textbf{0.9882} &  \textbf{0.067} \\
        & 120 & \textbf{42.80} & \textbf{0.9904} &  \textbf{0.058} \\
        \bottomrule
    \end{tabular}
    \label{ablation_table}
\end{table}

\begin{table}[htbp]
    \centering
    \caption{Generalization Study of Random Masking \& Wavelet HF.}
    \vspace{-5pt}
    \begin{tabular}{c|c|ccc}
        \toprule
        Methods & Views & PSNR(dB) & SSIM & MSE($10^{-3}$) \\
        \midrule
        \multirow{3}{*}{SHD+NMS} 
        & 60 & 38.06 & 0.9884 & 0.159 \\
        & 90 & 44.17 & 0.9956 & 0.039 \\
        & 120 & 46.41 & 0.9968 & 0.023 \\
        \cmidrule{1-5}
        \multirow{3}{*}{SRM+NRH} 
        & 60 & 37.05 & 0.9845 & 0.200 \\
        & 90 & 42.84 & 0.9938 & 0.053 \\
        & 120 & 45.57 & 0.9959 & 0.028 \\
        \cmidrule{1-5} 
        \multirow{3}{*}{SWARM} 
        & 60 &  \textbf{38.38} &  \textbf{0.9887} &  \textbf{0.146} \\
        & 90 & \textbf{44.73} & \textbf{0.9958} &  \textbf{0.034} \\
        & 120 & \textbf{46.80} & \textbf{0.9969} & \textbf{0.021} \\
        \bottomrule
    \end{tabular}
    \label{generalization_table}
\end{table}

\section{Discussion and Conclusion}

In this study, we introduce a diffusion strategy for sparse-view computed tomography (SVCT) reconstruction that integrates random masking and high-frequency wavelet decomposition. The proposed method significantly enhances model performance in terms of global consistency and detail reconstruction, thereby improving the accuracy and reliability of reconstructed images. Testing across multiple datasets demonstrates that the SWARM approach exhibits strong generalization and robustness. However, the method faces challenges stemming from the inherent nature of diffusion models, which result in lengthy computation times for full sampling steps. Future research will explore optimizing the diffusion process via skip sampling to reduce computational complexity and enhance sampling efficiency while maintaining high model performance. Additionally, we plan to investigate mask designs compatible with CT scanning geometry and dynamically adaptive masks that incorporate anatomical features to further enhance the practicality and robustness of this innovative approach.

\section*{References}

\bibliographystyle{IEEEtran}
\bibliography{IEEEabrv,reference}

@article{cormack1963representation,
  title={Representation of a function by its line integrals, with some radiological applications},
  author={Cormack, Allan Macleod},
  journal={J. Appl. Phys.},
  volume={34},
  number={9},
  pages={2722-2727},
  year={1963},
  publisher={AIP Publishing}
}

@article{hounsfield1973computerized,
  title={Computerized transverse axial scanning (tomography): Part 1. Description of system},
  author={Hounsfield, Godfrey N},
  journal={Br J Radiol},
  volume={46},
  number={552},
  pages={1016-1022},
  year={1973},
  publisher={The British Institute of Radiology}
}

@article{Wang2024A,
title={A Review of Deep Learning CT Reconstruction From Incomplete Projection Data},
author={Wang, Tao and Xia, Wenjun and others},
journal={IEEE Trans. Radiat. Plasma Med. Sci.},
volume={8},
number={2},
pages={138-152},
year={2024}
}

@article{Li2023Sparse-view,
title={Sparse-view CT Reconstruction with 3D Gaussian Volumetric Representation},
author={Li, Yingtai and Fu, Xueming and others},
journal={CoRR},
volume={abs/2312.15676},
year={2023}
}

@article{dempster1977maximum,
  title={Maximum likelihood from incomplete data via the EM algorithm},
  author={Dempster, Arthur P and others},
  journal={J. R. Stat. Soc.},
  volume={39},
  number={1},
  pages={1-22},
  year={1977},
  publisher={Wiley Online Library}
}

@article{donoho2006compressed,
  title={Compressed sensing},
  author={Donoho, David L},
  journal={IEEE Trans. Inf. Theory},
  volume={52},
  number={4},
  pages={1289-1306},
  year={2006},
  publisher={IEEE}
}

@article{Sidky2008Image,
title={Image Reconstruction in Circular Cone-Beam Computed Tomography by Constrained, Total-Variation Minimization},
author={Emil Y, Sidky and Pan, Xiaochuan and others},
journal={Phys. Med. Biol.},
volume={53},
number={17},
pages={4777-4807},
year={2008}
}

@article{Yu2009Compressed,
title={Compressed Sensing Based Interior Tomography},
author={Yu, Hengyong and Wang, Ge},
journal={Phys. Med. Biol.},
volume={54},
number={9},
pages={2791-2805},
year={2009}
}

@article{rioul1991wavelets,
  title={Wavelets and signal processing},
  author={Rioul, Olivier and Vetterli, Martin},
  journal={IEEE Trans. Inf. Theory},
  volume={8},
  number={4},
  pages={14-38},
  year={1991},
  publisher={IEEE}
}

@article{Han2016Deep,
title={Deep Residual Learning for Compressed Sensing CT Reconstruction Via Persistent Homology Analysis},
author={Yo Seob Han and Jaejun Yoo and others},
journal={CoRR},
volume={abs/1611.06391},
year={2016}
}

@article{Zhang2018A,
title={A Sparse-View CT Reconstruction Method Based on Combination of DenseNet and Deconvolution.},
author={Zhicheng Zhang and Xiaokun Liang and others},
journal={IEEE Trans. Med. Imaging},
volume={37},
number={6},
pages={1407-1417},
year={2018}
}

@article{Koetzier2023Deep,
title={Deep Learning Image Reconstruction for CT: Technical Principles and Clinical Prospects.},
author={Lennart R. Koetzier and Domenico Mastrodicasa and others},
journal={RADIOLOGY},
volume={306},
number={3},
year={2023}
}

@article{Zhong2024Impacts,
title={Impacts of Adaptive Statistical Iterative Reconstruction-V and Deep Learning Image Reconstruction Algorithms on Robustness of CT Radiomics Features: Opportunity for Minimizing Radiomics Variability among Scans of Different Dose Levels},
author={Jingyu Zhong and Zhiyuan Wu and others},
journal={J. Imaging Inform. Med.},
pages={1-11},
year={2024}
}

@article{McCann2017Deep,
title={Deep Convolutional Neural Network for Inverse Problems in Imaging},
author={Michael T. McCann and Kyong Hwan Jin and others},
journal={IEEE Trans. Image Process.},
volume={26},
number={9},
pages={4509-4522},
year={2017}
}

@article{Chen2017Low-Dose,
title={Low-Dose CT with a Residual Encoder-Decoder Convolutional Neural Network},
author={Hu Chen and Yi Zhang and others},
journal={IEEE Trans. Med. Imaging},
volume={36},
number={12},
pages={2524-2535},
year={2017}
}

@article{Zhang2020Artifact,
title={Artifact Removal Using a Hybrid-Domain Convolutional Neural Network for Limited-Angle Computed Tomography Imaging},
author={Qiyang Zhang and Zhanli Hu and others},
journal={Phys. Med. Biol.},
volume={65},
number={15},
pages={155010},
year={2020}
}

@article{Pan2022Multi-domain,
title={Multi-domain Integrative Swin Transformer Network for Sparse-View Tomographic Reconstruction},
author={Jiayi Pan and Heye Zhang and others},
journal={PATTERNS},
volume={3},
number={6},
year={2022}
}

@article{Hu2021Hybrid-Domain,
title={Hybrid-Domain Neural Network Processing for Sparse-View CT Reconstruction},
author={Dianlin Hu and Jin Liu and others},
journal={IEEE T. Radiat. Plasma Med. Sci.},
volume={5},
number={1},
pages={88-98},
year={2021}
}

@article{Xia2022Patch-Based,
title={Patch-Based Denoising Diffusion Probabilistic Model for Sparse-View CT
  Reconstruction},
author={Wenjun Xia and Wenxiang Cong and others},
journal={CoRR},
year={2022}
}

@article{Wu2023Data-iterative,
title={Data-iterative Optimization Score Model for Stable Ultra-Sparse-View CT
  Reconstruction},
author={Weiwen Wu and Yanyang Wang},
journal={CoRR},
volume={abs/2308.14437},
year={2023}
}

@article{Guan2023Generative,
title={Generative Modeling in Sinogram Domain for Sparse-view CT Reconstruction},
author={Bing Guan and Cailian Yang and others},
journal={IEEE T. Radiat. Plasma Med. Sci.},
volume={8},
number={2},
pages={195-207},
year={2023}
}

@misc{Xia2023Parallel,
title={Parallel Diffusion Model-based Sparse-view Cone-beam Breast CT},
author={Wenjun Xia and Hsin Wu Tseng and others},
year={2024},
      eprint={2303.12861},
      archivePrefix={arXiv},
      url={https://arxiv.org/abs/2303.12861}, 
}

@article{Yang2024A,
title={A Dual-domain Diffusion Model for Sparse-view CT Reconstruction},
author={Chun Yang and Dian Sheng and others},
journal={IEEE Trans. Image Process.},
volume={31},
pages={1279-1283},
year={2024}
}

@misc{kazerouni2023diffusionmodelsmedicalimage,
      title={Diffusion Models for Medical Image Analysis: A Comprehensive Survey}, 
      author={Amirhossein Kazerouni and Ehsan Khodapanah Aghdam and others},
      year={2023},
      eprint={2211.07804},
      archivePrefix={arXiv},
      url={https://arxiv.org/abs/2211.07804}, 
}

@misc{ho2020denoisingdiffusionprobabilisticmodels,
      title={Denoising Diffusion Probabilistic Models}, 
      author={Jonathan Ho and Ajay Jain and others},
      year={2020},
      eprint={2006.11239},
      archivePrefix={arXiv},
      url={https://arxiv.org/abs/2006.11239}, 
}

@misc{toker2024satsynthaugmentingimagemaskpairs,
      title={SatSynth: Augmenting Image-Mask Pairs through Diffusion Models for Aerial Semantic Segmentation}, 
      author={Aysim Toker and Marvin Eisenberger and others},
      year={2024},
      eprint={2403.16605},
      archivePrefix={arXiv},
      url={https://arxiv.org/abs/2403.16605}, 
}

@misc{konz2024anatomicallycontrollablemedicalimagegeneration,
      title={Anatomically-Controllable Medical Image Generation with Segmentation-Guided Diffusion Models}, 
      author={Nicholas Konz and Yuwen Chen and others},
      year={2024},
      eprint={2402.05210},
      archivePrefix={arXiv},
      url={https://arxiv.org/abs/2402.05210}, 
}

@article{tan2024msdiff,
  title={MSDiff: Multi-Scale Diffusion Model for Ultra-Sparse View CT Reconstruction},
  author={Pinhuang Tan and  Mengxiao Geng and others},
  journal={arXiv preprint arXiv:2405.05814},
  year={2024}
}

@article{song2019generative,
  title={Generative modeling by estimating gradients of the data distribution},
  author={Song, Yang and Ermon, Stefano},
  journal={Adv. Neural Inf. Process. Syst.},
  volume={32},
  pages={11895-11907},
  year={2019}
}

@article{wang2024flame,
  title={FLAME Diffuser: Wildfire Image Synthesis using Mask Guided Diffusion},
  author={Wang, Hao and Boroujeni, Sayed Pedram Haeri and others},
  journal={arXiv preprint arXiv:2403.03463},
  year={2024}
}

@inproceedings{gao2023masked,
  title={Masked diffusion transformer is a strong image synthesizer},
  author={Gao, Shanghua and Zhou, Pan and others},
  booktitle={Proc. IEEE/CVF Int. Conf. Comput. Vis.},
  pages={23164-23173},
  year={2023}
}

@article{aversa2024diffinfinite,
  title={Diffinfinite: Large mask-image synthesis via parallel random patch diffusion in histopathology},
  author={Aversa, Marco and Nobis, Gabriel and others},
  journal={NeurIPS},
  volume={36},
  pages={78126-78141},
  year={2024}
}

@article{couairon2022diffedit,
  title={Diffedit: Diffusion-based semantic image editing with mask guidance},
  author={Couairon, Guillaume and Verbeek, Jakob and others},
  journal={arXiv preprint arXiv:2210.11427},
  year={2022}
}

@article{wang2023instructedit,
  title={Instructedit: Improving automatic masks for diffusion-based image editing with user instructions},
  author={Wang, Qian and Zhang, Biao and others},
  journal={arXiv preprint arXiv:2305.18047},
  year={2023}
}

@inproceedings{zou2024towards,
  title={Towards Efficient Diffusion-Based Image Editing with Instant Attention Masks},
  author={Zou, Siyu and Tang, Jiji and others},
  booktitle={AAAI Conf. Artif. Intell.},
  volume={38},
  number={7},
  pages={7864-7872},
  year={2024}
}

@article{pang2024improved,
  title={An Improved Face Image Restoration Method Based on Denoising Diffusion Probabilistic Models},
  author={Pang, Yun and Mao, Jiawei and others},
  journal={IEEE Access},
  pages={3581-3596},
  year={2024}
}

@inproceedings{zhu2023denoising,
  title={Denoising diffusion models for plug-and-play image restoration},
  author={Zhu, Yuanzhi and Zhang, Kai and others},
  booktitle={Proc. IEEE/CVF Conf. Comput. Vis. Pattern Recognit.},
  pages={1219-1229},
  year={2023}
}

@inproceedings{le2024maskdiff,
  title={Maskdiff: Modeling mask distribution with diffusion probabilistic model for few-shot instance segmentation},
  author={Le, Minh-Quan and Nguyen, Tam V and others},
  booktitle={AAAI Conf.},
  volume={38},
  number={3},
  pages={2874-2881},
  year={2024}
}

@inproceedings{toker2024satsynth,
  title={Satsynth: Augmenting image-mask pairs through diffusion models for aerial semantic segmentation},
  author={Toker, Aysim and Eisenberger, Marvin and others},
  booktitle={CVPR},
  pages={27695-27705},
  year={2024}
}

@article{song2021solving,
  title={Solving inverse problems in medical imaging with score-based generative models},
  author={Song, Yang and Shen, Liyue and others},
  journal={arXiv preprint arXiv:2111.08005},
  year={2021}
}

@article{rajpurkar2022ai,
  title={AI in health and medicine},
  author={Rajpurkar, Pranav and Chen, Emma and others},
  journal={Nat. Med.},
  volume={28},
  number={1},
  pages={31-38},
  year={2022}
}

@article{song2020score,
  title={Score-based generative modeling through stochastic differential equations},
  author={Song, Yang and Sohl-Dickstein, Jascha and others},
  journal={arXiv preprint arXiv:2011.13456},
  year={2020}
}

@article{brenner2007computed,
  title={Computed tomography—an increasing source of radiation exposure},
  author={Brenner, David J and Hall, Eric J},
  journal={N. Engl. J. Med.},
  volume={357},
  number={22},
  pages={2277-2284},
  year={2007}
}

@article{AAPM,
  title={Low dose CT grand challenge.},
    journal={Available: http://www.aapm.org/GrandChallenge/LowDoseCT/},
  year={2017}
}

@article{article1,
title = {An explainable deep-learning algorithm for the detection of acute intracranial haemorrhage from small datasets},
author = {Lee, Hyunkwang and Yune, Sehyo and others},
journal = {Nat. Biomed. Eng.},
volume = {3},
pages = {173-182},
year = {2019}
}

@article{article,
title = {Artificial intelligence in cancer imaging: Clinical challenges and applications},
author = {Bi, Wenya Linda and Hosny, Ahmed and others},
journal = {CA Cancer J Clin},
month = {02},
volume = {69},
pages = {127-157},
year = {2019}
}

@article{ABDAR2021243,
title = {A review of uncertainty quantification in deep learning: Techniques, applications and challenges},
author = {Moloud, Abdar and Farhad, Pourpanah and others},
journal = {Information Fusion},
volume = {76},
pages = {243-297},
year = {2021}
}

@article{10403850,
  author={Xu, Kai and Lu, Shiyu and Huang, Bin and Wu, Weiwen and Liu, Qiegen},
  journal={IEEE Transactions on Medical Imaging}, 
  title={Stage-by-Stage Wavelet Optimization Refinement Diffusion Model for Sparse-View CT Reconstruction}, 
  volume={43},
  number={10},
  pages={3412-3424},
  year={2024}
  }

@article{acar2024advanced,
  title={Advanced hyperthermia treatment: optimizing microwave energy focus for breast cancer therapy},
  author={Acar, Burak and YILMAZ ABDOLSAHEB, TUBA and Yapar, Ali},
  journal={Turkish Journal of Electrical Engineering and Computer Sciences},
  volume={32},
  number={2},
  pages={268-284},
  year={2024}
}

@article{ozturk2024diffusion,
  title={Diffusion Probabilistic Models for Image Formation in MRI},
  author={{\"O}zt{\"u}rk, {\c{S}}aban and G{\"u}ng{\"o}r, Alper and {\c{C}}ukur, Tolga},
  booktitle={Generative Machine Learning Models in Medical Image Computing},
  pages={341-360},
  year={2024},
  publisher={Springer}
}

@article{ozbey2023unsupervised,
  title={Unsupervised medical image translation with adversarial diffusion models},
  author={{\"O}zbey, Muzaffer and Dalmaz, Onat and Dar, Salman UH and Bedel, Hasan A and {\"O}zturk, {\c{S}}aban and G{\"u}ng{\"o}r, Alper and {\c{C}}ukur, Tolga},
  journal={IEEE Transactions on Medical Imaging},
  volume={42},
  number={12},
  pages={3524-3539},
  year={2023},
  publisher={IEEE}
}

@article{gungor2023adaptive,
  title={Adaptive diffusion priors for accelerated MRI reconstruction},
  author={G{\"u}ng{\"o}r, Alper and Dar, Salman UH and {\"O}zt{\"u}rk, {\c{S}}aban and Korkmaz, Yilmaz and Bedel, Hasan A and Elmas, Gokberk and Ozbey, Muzaffer and {\c{C}}ukur, Tolga},
  journal={Medical image analysis},
  volume={88},
  pages={102872},
  year={2023},
  publisher={Elsevier}
}

@article{HUANG2025103334,
title = {Enhancing global sensitivity and uncertainty quantification in medical image reconstruction with Monte Carlo arbitrary-masked mamba},
author = {Jiahao, Huang and Liutao, Yang and others},
journal = {Medical Image Analysis},
volume = {99},
pages = {103334},
year = {2025},
issn = {1361-8415}
}

\clearpage

\appendices
 
 \section{Performance Comparison of  Ultra-Sparse Views} 

\begin{table}[htbp]
    \centering
    \caption{Performance Comparison of Different Methods Under Ultra-Sparse Views on AAPM Challenge Data.}  
    \begin{tabular}{l|c|ccc}  
        \toprule
       Methods & Views & PSNR(dB) & SSIM & MSE($10^{-3}$) \\
        \midrule
        \multirow{3}{*}{GMSD} 
        & 15 & 26.69 & 0.8741 & 2.318 \\
        & 30 & 31.11 & 0.9363 & 0.869 \\
        & 45 & 34.76 & 0.9598 & 0.368 \\
        \cmidrule{1-5}  
        \multirow{3}{*}{SWORD} 
        & 15 & \textbf{27.29} & \textbf{0.8833} & \textbf{1.991} \\
        & 30 & 33.11 & 0.9492 & 0.529 \\
        & 45 & 36.96 & 0.9728 & 0.211 \\
        \cmidrule{1-5} 
        \multirow{3}{*}{SWARM} 
        & 15 & 27.12 & 0.8733 & 2.414 \\  
        & 30 & \textbf{34.12} & \textbf{0.9627} & \textbf{0.438} \\
        & 45 & \textbf{37.30} & \textbf{0.9768} & \textbf{0.213} \\
        \bottomrule
    \end{tabular}
    \label{other_sparse_view}  
\end{table}

To evaluate the limit performance of the proposed method, we further conducted experiments under ultra-sparse view conditions. As shown in Table \ref{other_sparse_view}, our method achieves superior reconstruction performance at 30 and 45 views, while the SWORD method performs slightly better under the extremely sparse condition of 15 views. Nevertheless, our method maintains the best overall performance across most view settings, fully demonstrating its effectiveness and stability.



\section{Proof of Proposition 3.1}
\label{appA}

\textbf{Proof:} Let $y_{M}$ represent the sinogram after random masking $m$. The perturbed data $\tilde{y}$ can be expressed as the sum of the sinogram and its corresponding mask, such that $\tilde{y}=y+y_{M}$.  The mean of the perturbed data $\tilde{y}$, denoted as $\tilde{\mu}$, can be formulated as follows:
\begin{equation}
\label{eq4}
\tilde{\mu}= \frac{1}{n}\Sigma^{n}_{i=1}(y_{i}+y_{M_{i}}).  
\end{equation}
The variance of the data after the application of masking can be represented as follows:
\begin{equation}
\label{eq5}
\sigma^{2}(\tilde{y})=\frac{1}{n} \Sigma^{n}_{i=1}((y_{i}-\mu)+(y_{M_{i}}-\mu_{M}))^{2},
\end{equation}
where $\sigma^{2}(\tilde{y})$  is the variance of $\tilde{y}$, $\mu$ and $\mu_{M}$ represent the mean of $y$ and $y_{M}$, respectively. Depending on the properties of expectation and variance, Equation  \eqref{eq5}  can be rewritten as:
 \begin{equation}
 \label{eq6}
\begin{split}
\sigma^{2}(\tilde{y}) =  \sigma^{2}(y) + \frac{1}{n} \Sigma^{n}_{i=1} 2(y_{i}-\mu)(y_{M_{i}}-\mu_{M})    \\
+\frac{1}{n} \Sigma^{n}_{i=1} (y_{M_{i}}-\mu_{M})^{2},
\end{split}
\end{equation}
where $\sigma^{2}(y)$  is the variance of $y$. For an arbitrary random mask $m_{i}$, we have $y_{M_{i}}=y_{i} \cdot m_{i}$ and $\mu_{M}=\frac{1}{n}\Sigma^{n}_{j=1} y_{j}\cdot m_{j} $. Extracting the cross terms of Equation \eqref{eq6}, then
 \begin{equation}
\begin{split}
\mathcal{F}  &=  \frac{1}{n} \Sigma^{n}_{i=1} 2(y_{i}-\mu)(y_{M_{i}}-\mu_{M})   \\
&= \frac{1}{n}\Sigma^{n}_{i=1}2(y_{i}-\mu)(y_{i}m_{i}-\frac{1}{n}\Sigma^{n}_{j=1}y_{j}m_{j}).
\end{split}
  \label{eq7}
\end{equation}
\setlength{\baselineskip}{13pt}
$\mathcal{F}$ can be decomposed into the sum of two terms, i.e., $\mathcal{F}=\mathcal{F}_{1}-\mathcal{F}_{2}=\frac{1}{n}\Sigma^{n}_{i=1}2(y_{i}-\mu)(y_{i}m_{i})-\frac{1}{n}\Sigma^{n}_{i=1}2(y_{i}-\mu)(\frac{1}{n}\Sigma^{n}_{j=1}y_{j}m_{j})$.  It implies that $\mathbb{E}[m_{i}]=\frac{a+b}{2}$ since $m_{i}\sim \mathnormal{U}(a,b)$.  Let $\mathbb{E}[\cdot]$ stands for expectation. Then we have $\mathbb{E}[\mathcal{F}_{1}]=\frac{1}{n}\Sigma^{n}_{i=1}2(y_{i}-\mu)y_{i}\mathbb{E}[m_{i}]=\frac{a+b}{2}\cdot \frac{2}{n} \Sigma^{n}_{i=1}(y_{i}-\mu)y_{i} \geq 0 $ since $\Sigma^{n}_{i=1}(y_{i}-\mu)y_{i}=\sigma^{2}(y)$ is the variance term, the expectation is $0$. Moreover,  $\mathbb{E}[\mathcal{F}_{2}]=-\frac{2}{n^{2}}\Sigma^{n}_{i=1} \Sigma^{n}_{j=1}(y_{i}-\mu) \cdot y_{j} \cdot \frac{a+b}{2}=0$ since $\Sigma^{n}_{i=1}(y_{i}-\mu)=0$. Hence, $\mathbb{E}[\mathcal{F}]=0$.

Since $\mathbb{E}[\frac{1}{n}\Sigma^{n}_{i=1}(y_{M_{i}}-\mu_{M})^{2}] \geq 0$, it means that $\mathbb{E}[\sigma^{2}(\tilde{y})] \geq \sigma^{2}(y) $.    {\rsq}

\section{Proof of Proposition 3.2}
\label{appB}

\textbf{Proof:} Let the original data sample space be $ y = \{ y_1, y_2, \cdots, y_n \} $, and the masked data sample space be $ \tilde{y} = y + y_M = y + m \odot y $, where $ m $ represents the random mask. Then, the expectation of the masked data:
\begin{equation}
\label{eq_ap1}
\tilde{\mu} = \mathbb{E}[\tilde{y}] = \mathbb{E}[y + m \odot y] = \mu + \mathbb{E}[m \odot y],  
\end{equation}
 where $ \mu $ and $\tilde{\mu}$ represent the means of the original data sample and the masked data sample, respectively. Therefore, the deviation of the masked data from its expectation is: 
 \begin{equation}
 \label{eq_ap2}
\begin{split}
\tilde{y} - \tilde{\mu} &=  (y + m \odot y) - (\mu + \mathbb{E}[m \odot y])   \\
&= (y - \mu) + (m \odot y - \mathbb{E}[m \odot y]).
\end{split}
\end{equation}
 The covariance of the perturbed sample space is as follows:
\begin{equation}
\label{eq_ap3}
\begin{split}
\tilde{\Sigma}  &=  \mathbb{E}\left[ (\tilde{y} - \tilde{\mu})(\tilde{y} - \tilde{\mu})^{T} \right]   \\
& = \mathbb{E}\left[ (y - \mu)(y - \mu)^{T} \right] 
    + \mathbb{E}\left[ (y - \mu)\left( m \odot y - \mathbb{E}[m \odot y] \right)^{T} \right]   \\
&  + \mathbb{E}\left[ \left( m \odot y - \mathbb{E}[m \odot y] \right)(y - \mu)^{T} \right]  \\
&    + \mathbb{E}\left[ \left( m \odot y - \mathbb{E}[m \odot y] \right)\left( m \odot y - \mathbb{E}[m \odot y] \right)^{T} \right].
\end{split}
\end{equation}
Let \( \mathcal{G} \) be the total covariance: \( \mathcal{G} = \mathcal{G}_1 + \mathcal{G}_2 + \mathcal{G}_3 + \mathcal{G}_4 \) and $\mathcal{G}_1 = \mathbb{E}\left[ (y - \mu)(y - \mu)^{T} \right] = \Sigma$, where $ \Sigma = \mathbb{E}\left[ (y - \mu)(y - \mu)^{T} \right]$. Since $ \mu $ and $ y $ are independent and the mask space follows a uniform distribution, then,
\begin{equation}
\label{eq_ap4}
\begin{split}
\mathcal{G}_2  &=  \mathbb{E}\left[ (y-\mu)\left( m \odot y - \mathbb{E}[m \odot y] \right)^{T} \right]   \\
& =  \mathbb{E}\left[ (y-\mu)\left( (m-\mathbb{E}[m])\odot \mu + m \odot (y-\mu) \right)^{T} \right]   \\
& = \mathbb{E}[m] \odot \mathbb{E}\left[ (y-\mu)(y-\mu)^{T} \right]  \\
& = \mathbb{E}[m] \odot \Sigma \geq 0.
\end{split}
\end{equation}
 Since $ \Sigma $ is positive semi-definite and $ \mathbb{E}\left[m\right] \geq 0 $, it follows that $ \mathcal{G}_{2} \geq 0 $. Similarly, it can be obtained that $ \mathcal{G}_{2} = \mathcal{G}_{3}^{T} = \mathbb{E}\left[ \left( m \odot y - \mathbb{E}[m \odot y] \right)(y - \mu)^{T} \right] \geq 0 $.  Furthermore, let $ y_M = m \odot y $, then $ \mathcal{G}_4  = \mathbb{E}\left[ (y_M - \mathbb{E}[y_M])(y_M - \mathbb{E} [ y_M ])^T \right] $, which is the covariance matrix of $ y_M $. Hence, $ \mathcal{G}_4 $ is positive semi-definite  and it follows that $ \mathcal{G}_4 \succeq 0 $.  Therefore, $ \tilde{\Sigma} = \Sigma + \varepsilon $, where $ \varepsilon = \mathcal{G}_2 +\mathcal{G}_3 +\mathcal{G}_4 \geq 0 $ represents the covariance perturbation increment. Then, $\tilde{\Sigma} \geq \Sigma$.  For any direction $ \nu\in \mathbb{R}^{d} $, it holds that $ \nu^{T} \tilde{\Sigma} \nu \geq \nu^{T} \Sigma \nu $. This implies that the variance of the perturbed data does not decrease along every direction $\nu $.
 
As the data dispersion increases, the data samples are no longer confined to their original clustered regions but instead expand along specific directions into a broader feature space, thereby extending the boundaries of the data distribution. Experimental results further validate the significant effectiveness of random masks in expanding the coverage of the data distribution.   {\rsq}

\end{document}